%% file: main.tex
\documentclass{sig-alternate}
\usepackage{textcomp}
\usepackage{textcmds}
\usepackage{mathptmx} 
\usepackage{multirow}
\usepackage{fancyhdr}
\usepackage[normalem]{ulem}
\usepackage[hyphens]{url}
\usepackage[sort,nocompress]{cite}
\usepackage[final]{microtype}
\usepackage{flushend}
\usepackage[ruled,linesnumbered]{algorithm2e}
\usepackage{listings}
\usepackage{xcolor}
\usepackage{footnote}
\usepackage{threeparttable}
\usepackage{paralist}
\usepackage{cite}
\usepackage{booktabs}
\usepackage{authblk}

\usepackage[bookmarks=true,breaklinks=true,letterpaper=true,colorlinks,linkcolor=black,citecolor=blue,urlcolor=black]{hyperref}

\usepackage{multirow}
\newcommand{\squishlist}{
   \begin{list}{$\bullet$}
    { \setlength{\itemsep}{0pt}      \setlength{\parsep}{0pt}
      \setlength{\topsep}{3pt}       \setlength{\partopsep}{0pt}
      \setlength{\listparindent}{-2pt}
      \setlength{\itemindent}{-5pt}
      \setlength{\leftmargin}{1em} \setlength{\labelwidth}{0em}
      \setlength{\labelsep}{0.5em} } }

\newcommand{\squishend}{
    \end{list}  }

\fancypagestyle{firstpage}{
     \fancyhf{}
     \setlength{\headheight}{10pt}
     
     \fancyhead[C]{To Appear in 2020 IEEE International Symposium on High Performance Computer Architecture \vspace{-20pt}}
     \pagenumbering{arabic}
}

\title{HyGCN: A GCN Accelerator with Hybrid Architecture \vspace{-40pt}}

\author[$^{\dag}$ $^{\ddag}$ $^{\S}$]{\textbf{Mingyu Yan}}
\author[$^{\S}$]{\textbf{Lei Deng}}
\author[$^{\S}$]{\textbf{Xing Hu}}
\author[$^{\S}$]{\textbf{Ling Liang}}
\author[$^{\dag}$]{\textbf{Yujing Feng}}
\author[$^{\dag}$ *]{\\ \textbf{Xiaochun Ye}}
\author[$^{\dag}$]{\textbf{Zhimin Zhang}}
\author[$^{\dag}$ $^{\ddag}$]{\textbf{Dongrui Fan}}
\author[$^{\S}$]{\textbf{Yuan Xie} \vspace{-10pt}}

\affil[$^{\dag}$]{State Key Laboratory of Computer Architecture, Institute of Computing Technology, Chinese Academy of Sciences} 
\affil[$^{\ddag}$]{School of Computer Science and Technology, University of Chinese Academy of Sciences}
\affil[$^{\S}$]{University of California, Santa Barbara \vspace{50pt}}

\begin{document}

\maketitle

\thispagestyle{firstpage}
\pagestyle{plain}
\footnotetext{$^*$Corresponding author is Xiaochun Ye and his email is yexiaochun@ict.ac.cn.}

\begin{abstract}
Inspired by the great success of neural networks, graph convolutional neural networks (GCNs) are proposed to analyze graph data.
GCNs mainly include two phases with distinct execution patterns. 
The \emph{Aggregation} phase, behaves as graph processing, showing a dynamic and irregular execution pattern. The \emph{Combination} phase, acts more like the neural networks, presenting a static and regular execution pattern. 
The hybrid execution patterns of GCNs require a design that alleviates irregularity and exploits regularity.
Moreover, to achieve higher performance and energy efficiency, the design needs to leverage the high intra-vertex parallelism in \textit{Aggregation} phase, the highly reusable inter-vertex data in \textit{Combination} phase, and the opportunity to fuse phase-by-phase execution introduced by the new features of GCNs. 
However, existing architectures fail to address these demands.

In this work, we first characterize the hybrid execution patterns of GCNs on Intel Xeon CPU.
Guided by the characterization, we design a GCN accelerator, \textit{HyGCN}, using a hybrid architecture to efficiently perform GCNs.
Specifically, first, we build a new programming model to exploit the fine-grained parallelism for our hardware design.
Second, we propose a hardware design with two efficient processing engines to alleviate the irregularity of \emph{Aggregation} phase and leverage the regularity of \emph{Combination} phase. 
Besides, these engines can exploit various parallelism and reuse highly reusable data efficiently.
Third, we optimize the overall system via inter-engine pipeline for inter-phase fusion and priority-based off-chip memory access coordination to improve off-chip bandwidth utilization.
Compared to the state-of-the-art software framework running on Intel Xeon CPU and NVIDIA V100 GPU, our work achieves on average 1509$\times$ speedup with 2500$\times$ energy reduction and average 6.5$\times$ speedup with 10$\times$ energy reduction, respectively.
\end{abstract}

\input{tex/intro}

\input{tex/background}
\input{tex/motivation}

\input{tex/design.tex}

\input{tex/methodology.tex}

\input{tex/results.tex}
\input{tex/discussion}
\input{tex/related}

\input{tex/conclusion}

\section*{Acknowledgments}
We thank the anonymous reviewers of HPCA 2020 and the sealer in Scalable Energy-efficient Architecture Lab (SEAL) for their constructive and insightful comments. 
This work was supported by the National Key Research and Development Program of China (Grant No. 2018YFB1003501), the National Natural Science Foundation of China (Grant No. 61732018, 61872335, and 61802367), the Strategic Priority Research Program of Chinese Academy of Sciences (Grant No. XDA 18000000), the Innovation Project Program of the State Key Laboratory of Computer Architecture (Grant No. CARCH4408, CARCH4412, and CARCH4502), the National Science Foundation (Grant No. 1730309, 1725447 and CCF 1740352), and SRC nCORE NC-2766-A.

\bibliographystyle{IEEEtranS}
\bibliography{refs}

\end{document}

%% file: tex/intro.tex
\section{Introduction}
Inspired by the powerful learning capability of neural networks, graph convolutional neural networks (GCNs) are proposed as an effective category of models to represent and process graph data~\cite{AliGraph,comprehensive_gnn_survey,graph_deep_learning,GNN_Review}. GCNs convert the graph data into a low dimensional space while keeping both the structure and property information to the maximum extent, and then construct a neural network for the consequent training and inference. Recently, GCNs attract substantial efforts from both the industrial and academic communities \cite{GraphSage,web_scale_GCN,pytorch_biggraph,euler,neural_fps,DIFFPOOL,AS-GCN} to solve problems including node classification \cite{1stChebNet}, link prediction \cite{neural_fps,protein_interface_prediction}, graph clustering \cite{DIFFPOOL}, and recommendation \cite{SSE}. As a result, GCNs gradually become a new workload family member in data-centers, such as in Google~\cite{Graph_Nets_library}, Facebook \cite{pytorch_biggraph}, and Alibaba~\cite{euler,AliGraph}.  

The convolutional layers occupy the major execution time of GCNs through two primary execution phases: \emph{Aggregation} and \emph{Combination} \cite{AliGraph,Jumping_Knowledge_Networks,PyTorch_Geometric}. The \emph{Aggregation} phase maintains most graph processing behaviors. It heavily relies on the graph structure that is inherently random and sparse. Processing of each vertex requires aggregating features from all its source neighbours. Unfortunately, the amount and location of these source neighbors vary significantly among vertices. As a result, the computational graph~\cite{Computation_DAG} and memory access pattern in the \textit{Aggregation} phase of each vertex are dynamic and irregular.
The \emph{Combination} phase acts more like the neural networks. It transforms the feature vector of each vertex to a new one using a multi layer perceptron (MLP), which is usually expressed by a matrix-vector multiplication (MVM). Due to the identical connection pattern of each neuron within a neural network layer, the computational graph~\cite{Computation_DAG} and memory access pattern in the \textit{Combination} phase of each vertex are static and regular. 
Besides, there are additional characteristics in these two phases that distinguish GCNs from conventional workloads.
First, the length of vertex property is short and fixed in conventional graph analytics. However, in GCNs, the feature vector of each vertex is quite long and variable across layers, which introduces high-degree intra-vertex parallelism in \textit{Aggregation} phase. 
Second, the parameters in conventional MLP-based neural networks are never shared, while they can be fully shared among vertices in GCNs, which induces abundant highly reusable inter-vertex data in \textit{Combination} phase.
Third, the two phases are executed alternatively. An inherent dataflow exists between phases, providing an opportunity to fuse the phase-by-phase execution.

To achieve high-performance and energy-efficient acceleration of GCNs, aforementioned characteristics have imposed new requirements on architecture design.
First, not only can the GCN architecture alleviate the irregularity in \textit{Aggregation} phase, but it can also exploit the regularity in \textit{Combination} phase. 
Second, it needs to exploit the high-degree intra-vertex parallelism and highly reusable inter-vertex data. 
Third, it is able to efficiently fuse the execution of these two phases.

Unfortunately, existing architectures fail to implement GCN-specific characteristics.
For CPUs, although they can employ complex caching and prefetching techniques to offset the processor-memory disparity by exploiting the regular access pattern~\cite{Sequential_hardware_prefetching}, they fail to address the abundant dynamic and irregular data accesses in the \emph{Aggregation} phase since the irregularity harms the predictability of memory accesses~\cite{prefetcher_blue_geneQ}.
Besides, it is difficult to efficiently implement the reuse of the highly reusable parameter data between computing units in CPUs as like TPU~\cite{TPU} and Eyeriss~\cite{eyeriss}. Thus, the energy-hungry data accesses to cache introduce high energy consumption~\cite{eyeriss}.
For GPUs, although they are well optimized for neural networks, they lack the ability to alleviate irregularity in \textit{Aggregation} phase, which significantly hinders the performance improvement~\cite{GraphDynS,Tigr}. 
Furthermore, although they leverage the regularity in \textit{Combination} phase, the data copy and synchronization between threads for the parameter reuse are expensive. 
For graph analytics and neural network accelerators, they are only optimized to alleviate irregularity or exploit regularity, rather than both simultaneously. 
At last, all of them are short of the ability to efficiently fuse the execution of these two phases. 
In conclusion, existing architectures are not the ideal platforms to execute GCNs.

In this work, we first characterize the hybrid execution patterns of GCN workloads on Intel Xeon CPU. 
Next, guided by the characterization, we propose a GCN accelerator, \emph{HyGCN}, using a hybrid architecture to efficiently perform GCNs. 
Specially, we first propose a programming model to achieve the hardware transparency for programmers and exploit fine-grained parallelism. It abstracts GCNs as edge-centric aggregation for the \emph{Aggregation} phase and MVMs for the \emph{Combination} phase.
Second, we design \emph{HyGCN} with two efficient processing engines, \emph{Aggregation Engine} and \emph{Combination Engine}, to accelerate the \emph{Aggregation} and \emph{Combination} phases, respectively. 
In \emph{Aggregation Engine}, interval-shard graph partitioning and window sliding-shrinking methods are introduced to alleviate irregularity by increasing data reuse and decreasing unnecessary accesses for sparsity, respectively. 
Additionally, we implement a vertex-disperse processing method to exploit the edge parallelism and intra-vertex parallelism.
In \emph{Combination Engine}, to leverage the regularity, we build multi-granular systolic arrays to perform MVMs in parallel and reuse the shared parameters. Besides, they can be flexibly used either independently for lower latency or in combination for lower energy. 
Third, to improve the overall execution, on the basis of individual optimizations of these two phases, we build a fine-grained inter-engine pipeline to fuse the phase-by-phase execution and propose a priority-based memory access coordination for the off-chip data accesses between the two engines. 

\noindent 

To summarize, we list our contributions as follows: \par

\squishlist
    \item We study an emerging domain, GCNs, from a computer architecture perspective and show that hybrid execution patterns exist in GCNs. Specially, the \emph{Aggregation} phase in GCNs presents a dynamic and irregular execution pattern, while \emph{Combination} phase is static and regular. 

    \item We propose a GCN accelerator, \textit{HyGCN}, using a hybrid architecture to efficiently perform GCNs. 
    First, we build a programming model to enable our hardware design to exploit various parallelisms inherent in this domain. 
    Next, we propose a hardware design to tackle irregularity and leverage regularity with \emph{Aggregation Engine} and \emph{Combination Engine}, respectively. 
    
    \item We propose a flexible inter-engine pipeline and a priority-based memory access coordination to efficiently fuse the execution of \textit{Aggregation} phase and \textit{Combination} phase. 

    \item We implement our architecture design in RTL and evaluate it using a detailed microarchitectural simulation. We use four well-known GCN models on six popular graph datasets. 
    Compared to the state-of-the-art software framework \textit{PyTorch Geometric} \cite{PyTorch_Geometric} running on Intel Xeon CPU and NVIDIA V100 GPU, our work achieves on average 1509$\times$ speedup with 2500$\times$ energy reduction and 6.5$\times$ speedup with 10$\times$ energy reduction, respectively.
    
     \squishend

%% file: tex/background.tex
\section{Background}
\begin{figure*}[!hptb]
    \vspace{2pt}
    \centering
    \includegraphics[page=1, width=0.8\textwidth]{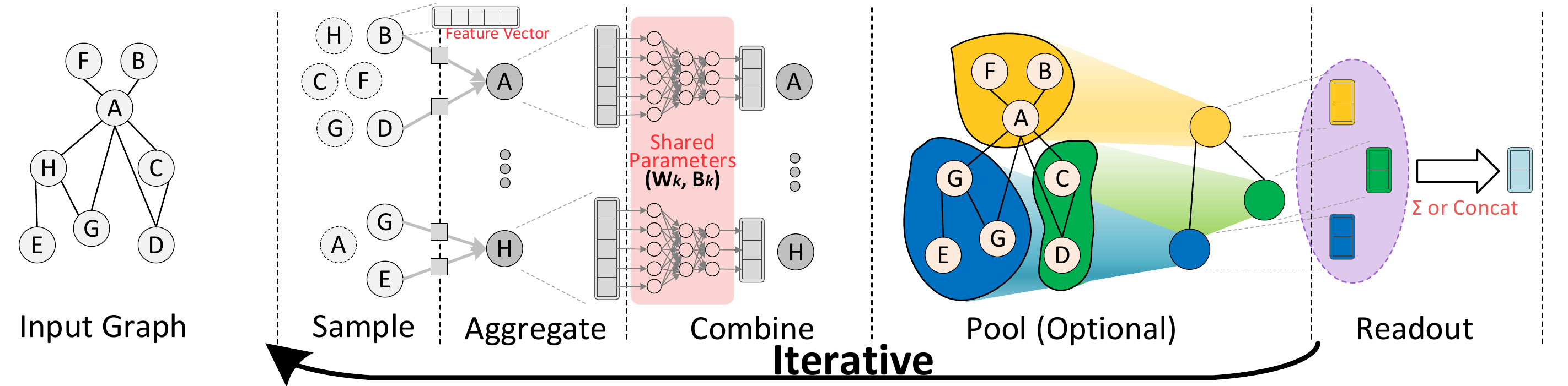}
    \vspace{-10pt}
    \caption{Illustration of the GCN model.}
    \label{fig:gcn_model}
     \vspace{-13pt}
\end{figure*}
GCNs follow a neighborhood aggregation scheme, where the feature vector of each vertex is computed by recursively aggregating and transforming the representation vectors of its neighbor vertices \cite{GINConv,GraphSage,AliGraph}. Fig. \ref{fig:gcn_model} illustrates the execution phases of GCN models. After $k$ iterations of aggregation via the \textbf{Aggregate} function and transformation via the \textbf{Combine} function, a vertex is represented by its final feature vector, which captures the structural information within the vertex's $k$-hop neighborhood. Table \ref{tab:notation} lists the notations used in GCNs. In this work, we mainly focus on undirected graphs and the inference stage rather than training.

\begin{table}[!hptb]
\vspace{-15pt}
\caption{GCN Notations.}\label{tab:notation}
\centering
\renewcommand\arraystretch{1.2}
 \resizebox{0.48\textwidth}{!}{
\begin{tabular}{cc|cc}
\toprule   
\textbf{Notation}    &\textbf{Meaning}   & \textbf{Notation}   &\textbf{Meaning}  \\ 
$\mathbf{G}$  & graph $\mathbf{G} = (V,E)$  &  $V$  & vertices of $\mathbf{G}$  \\
$E$  & edges of $\mathbf{G}$  & $D_v$ & degree of vertex $v$  \\
$e_{(i,j)}$  & edge between vertex $i$ and $j$  & $N(v)$ ($S(v)$) & (sampling subset of) $v$' neighbor set \\
$A$ ($A_{ij}$)  & (element of) adjacent matrix  & $a_v$ &  aggregation feature vector of $v$ \\
$h_G$ & feature vector of $\mathbf{G}$ & $W$ &  combination weight matrices \\
$h_v$  & feature vector of vertex $v$  & $b$ &  combination bias vectors \\
$X$  & initialized feature matrix  & $Z$ & embedding matrix \\
$C$  & assignment matrix  & $\epsilon$ &  learnable parameter \\
\bottomrule
\end{tabular}}
 \vspace{-10pt}
\end{table}

Typically, the $k$-th layer/iteration of GCNs is formulated as 
\begin{equation}
\setlength{\abovedisplayskip}{0pt}
\setlength{\belowdisplayskip}{0pt}
\begin{aligned}
a_v^k = \textbf{Aggregate} \big({h_u^{(k-1)}: u \in \{ N(v) \} \cup \{ v \}} \big), \\
h_v^k = \textbf{Combine} \big(a_v^k\big). 
\end{aligned}
\end{equation}
where $h_v^k$ is the representation feature vector of vertex $v$ at the $k$-th iteration. Simply, the \textbf{Aggregate} function aggregates multiple feature vectors from source neighbors to one single feature vector, and the \textbf{Combine} function transforms the feature vector of each vertex to another feature vector using an MLP neural network. Note that the MLP parameters, including weights and biases, are shared between vertices.

In order to decrease the computational complexity, the \textbf{Sample} function is usually applied before the \textbf{Aggregate} function to sample a subset from the neighbor vertices of each vertex \cite{GraphSage,FastGCN} as the new neighbors, specifically,
\begin{equation}
\setlength{\abovedisplayskip}{2pt}
\setlength{\belowdisplayskip}{2pt}
\begin{aligned}
S(v) = \textbf{Sample}^k \big(N(v) \big). \\
\end{aligned}
\end{equation}
Sometimes, the \textbf{Pool} function \cite{DIFFPOOL} follows the \textbf{Combine} function to transform the original graph into a smaller graph.

After several iterations, the features will be used for final prediction or classification. For the node classification problem, vertex feature vectors $h_v^k$ at the last iteration are used for prediction. For the graph classification problem, a \textbf{Readout} function further aggregates the $h_v^k$ at the last iteration to obtain the entire graph's representation vector, i.e.
\begin{equation}
\setlength{\abovedisplayskip}{2pt}
\setlength{\belowdisplayskip}{2pt}
h_G = \textbf{Readout} \big({h_v^k ~|~v \in G } \big).
\end{equation}

Next, we provide several typical GCN models as examples to explain the above operations in detail. 

\textbf{GCN} is one of the most successful convolutional networks for graph learning \cite{1stChebNet,comprehensive_gnn_survey}, which bridges the gap between spectral-based convolutions and spatial-based convolutions. Its inference model can be described as

\begin{equation}
\setlength{\abovedisplayskip}{1pt}
\setlength{\belowdisplayskip}{3pt}
\begin{aligned}
a_v^k = \big(\sum \frac{1}{\sqrt{D_v \cdot D_u}} h_{u}^{(k-1)} ~|~\forall u \in \{ N(v) \} \cup \{ v \}  \big ) ,\\
h_v^k = ReLU(W^k a_v^k + b^k). 
\end{aligned}
\end{equation}

\textbf{GraphSage} further adopts uniform neighbor sampling to alleviate receptive field expansion that effectively trades off accuracy and execution time \cite{GraphSage}. It is formulated as
\begin{equation}
\setlength{\abovedisplayskip}{3pt}
\setlength{\belowdisplayskip}{3pt}
\begin{aligned}
a_v^k = Mean \big(\{h_v^{(k-1)} \}\cup \{ h_u^{(k-1)} , \forall u \in S(v)  \} \big), \\
h_v^k = ReLU(W^k a_v^k + b^k). 
\end{aligned}
\end{equation}

\textbf{GINConv} is a simple neural architecture, and its discriminative power is equal to the power of the Weisfeiler-Lehman graph isomorphism test \cite{GINConv}. Vertex features learned by GINConv can be directly used for tasks like node classification and link prediction. We can perform this model as 
\begin{equation}
\setlength{\abovedisplayskip}{3pt}
\setlength{\belowdisplayskip}{3pt}
\begin{aligned}
a_v^k = (1+\epsilon_k) \cdot h_v^{(k-1)} + \sum_{u \in N(v)} h_u^{(k-1)}, \\
h_v^k = MLP^k(a_v^k,~W^k,~b^k).
\end{aligned}
\end{equation}

For graph classification tasks, the following \textbf{Readout} function is further used to produce the representation of the entire graph, given the representations of individual vertices. It concatenates across all iterations of GINConv to acquire the final graph representation as 
\begin{equation}
\setlength{\abovedisplayskip}{3pt}
\setlength{\belowdisplayskip}{3pt}
\begin{aligned}
h_G = \textbf{Concat} \big ( (\sum_{v \in G}{ h_v^k}) ~|~k = 1,...,K \big ).
\end{aligned}
\end{equation}

\textbf{DiffPool} provides a general tool to realize hierarchical graph-level transformation for a broad set of input graphs \cite{DIFFPOOL}. It can be inserted after the \textbf{Combine} function of any GCNs to transform the original graph to a smaller one (like the pooling layer in convolutional neural networks (CNNs)).
In fact, Diffpool uses two extra GCNs to implement the graph transformation, which follows
\begin{equation}
\setlength{\abovedisplayskip}{3pt}
\setlength{\belowdisplayskip}{3pt}
\label{equation_diffpool}
\begin{aligned}
C^{(k-1)} = softmax  \big (GCN^k_{pool}(A^{(k-1)},~X^{(k-1)})  \big), \\
Z^{(k-1)} = GCN^k_{embedding}(A^{(k-1)},~X^{(k-1)}), \\
X^{k} = C^{(k-1)^T} Z^{(k-1)},~~A^{k} = C^{(k-1)^T} A^{(k-1)} C^{(k-1)}.
\end{aligned}
\end{equation}
After the DiffPool transformation, a new feature matrix $X^{k}$ and adjacent matrix $A^{k}$ are produced, which can be combined to construct a new and smaller graph. In the new graph, $GCN^k_{pool}$ determines the number of vertices, and $GCN^k_{embedding}$ determines the length of vertex feature vector.

\textbf{Summary}. As explained above, we introduce several typical operations in GCNs: \emph{Sampling}, \emph{Aggregation}, \emph{Combination}, \emph{Pooling}, and \emph{Readout}. 
Except for \emph{Combination}, all the operations are graph structure-dependent, which involve graph processing. 
\emph{Combination} usually is a typical MLP neural network (single layer or multiple layers).
\emph{Sampling} is used to sample a subset from neighbors, which can be done during preprocessing \cite{AS-GCN} or with random selection during runtime \cite{GraphSage}. \emph{Aggregation} aggregates the features from its 1-hop neighbors. \emph{Pooling} acts like the pooling layer in CNNs to realize graph transformation by reducing the number of vertices and the length of feature vectors. \emph{Readout} can be a simple summation \cite{PyTorch_Geometric} across vertices or further concatenation across iterations \cite{GINConv}. Therefore, \emph{Readout} can be viewed as an extreme \emph{Aggregation}. This work  focuses on \emph{Aggregation} and \emph{Combination}, two major phases in GCNs.

%% file: tex/motivation.tex
\section{Motivation}
In this section, we quantitatively characterize and identify the hybrid execution patterns in processing GCNs. 
Next, we explain our motivation behind designing a GCN accelerator.

\subsection{Characterization on CPU}
\begin{figure}[!hptb] 
    \centering
    \vspace{-3pt}
    \includegraphics[page=1, width=0.66\linewidth]{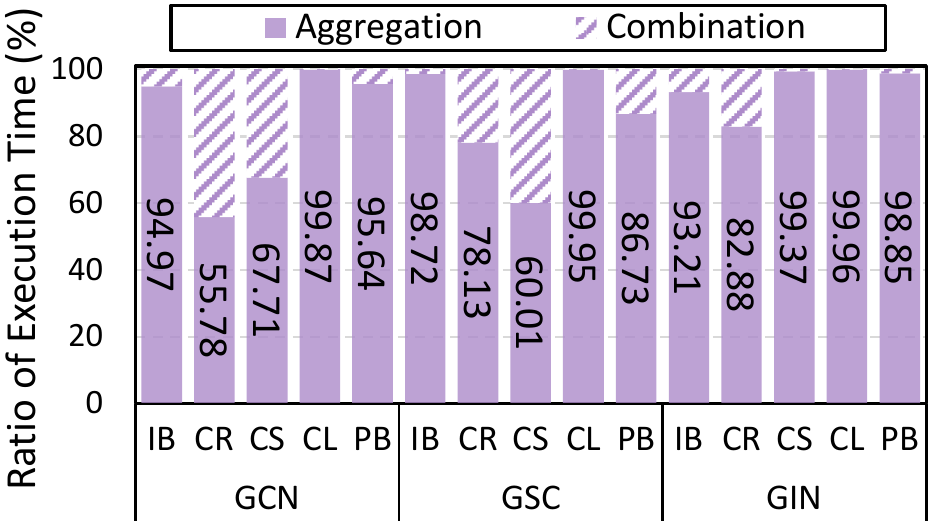}
    \vspace{-12pt}
    \caption{Execution time breakdown of the two phases.}
    \label{fig:execution_time_breakdown}
    \vspace{-10pt}
\end{figure}

We conduct quantitative characterizations using a state-of-the-art GCN software framework \emph{PyTorch Geometric} \cite{PyTorch_Geometric} on Intel Xeon CPU. The execution time breakdown of GCN (GCN)~\cite{1stChebNet}, GraphSage (GSC)~\cite{GraphSage}, and GINConv (GIN) ~\cite{GINConv} on several datasets~\cite{KKMMN2016} is illustrated in Fig.~\ref{fig:execution_time_breakdown}. The profiling results of GCN~\cite{1stChebNet} on the COLLAB dataset \cite{KKMMN2016} are presented in Table \ref{tab:characterization}. 
The details of system configuration and datasets are shown in Section~\ref{sec:setup}.

\noindent\textbf{Execution Time Breakdown.}
Both of \textit{Aggregation} and \textit{Combination} phases can occupy a significant amount of execution time, which implies that both need acceleration. Fig.~\ref{fig:execution_time_breakdown} illustrates their execution time ratio on different models and datasets. Their execution times differ due to the variable length of feature vectors and the execution flow of GCNs. For example, the long feature length of CR and CS datasets causes more time on \textit{Combination} phase for GCN and GraphSage. Since GINCov executes \textit{Aggregation} phase first, it spends more time on \textit{Aggregation} phase without the reduction of feature length through \textit{Combination} phase like the other two models do.

\noindent\textbf{Hybrid Execution Pattern.} 
The \emph{Aggregation} phase heavily relies on the graph structure that is inherently random and sparse, which results in numerous dynamic computations and irregular accesses.
From Table \ref{tab:characterization}, it is observed that each operation in the \emph{Aggregation} phase requires much more data to be accessed from DRAM than \textit{Combination} phase, resulting in higher DRAM access energy. 
Besides, the extremely high numbers of misses per kilo-instruction (MPKI) of L2 and L3 caches in the \emph{Aggregation} phase are caused by the high randomness of neighbor indices of each vertex. 
In addition, the indirect and irregular accesses render the data prefetching in the \emph{Aggregation} phase ineffective, since it is difficult to predict the data addresses without knowing the indices of neighbors in advance. 
This causes abundant ineffectual memory accesses to prefetch data. 

The \textit{Combination} phase executes a MVM for each vertex with a shared MLP-based neural network, which performs static and regular computations and accesses. 
Table \ref{tab:characterization} illustrates that each operation in the \emph{Combination} phase requires only small amount of data to be accessed from DRAM. This is because the MVMs are very compute-intensive and the weight matrix of MLP is widely shared between vertices. Nevertheless, up to 36\% of execution time for shared data copy and synchronization between threads is observed.

\begin{table}[!hptb]
 \vspace{-10pt}
\scriptsize
\centering
\caption{Quantitative Characterization on CPU.}\label{tab:characterization}
      \renewcommand\arraystretch{1.2}
      \resizebox{0.42\textwidth}{!}{
      \begin{tabular}{ccc}
      \toprule
      & \textbf{\emph{Aggregation}}         &  \textbf{\emph{Combination}}    \\ \midrule
      \textbf{DRAM Byte per Ops}            & 11.6                 & 0.06 \\  
      \textbf{DRAM Access Energy per Ops}   & 170$nJ$                 & 0.5$nJ$ \\
      \textbf{L2 Cache MPKI}                & 11                   &   1.5  \\        
      \textbf{L3 Cache MPKI}                & 10                   &   0.9  \\    
      \textbf{Ratio of Synchronization Time}    & ---              &   36\%  \\ 
      \bottomrule
      \end{tabular}}
\end{table}


\begin{table}[!hptb]
    \vspace{-20pt}
    \scriptsize
    \centering
    \caption{Different Execution Patterns of \emph{Aggregation} Phase and \emph{Combination} Phase.}\label{tab:variable_pattern}
	\renewcommand\arraystretch{1.2}
    \resizebox{0.45\textwidth}{!}{
    \begin{tabular}{ccc}
    \toprule
                                      & \textbf{\emph{Aggregation}}    &  \textbf{\emph{Combination}}      \\ \midrule
      \textbf{Access Pattern}         & Indirect \& Irregular          &   Direct \& Regular      \\         
      \textbf{Data Reusability}       & Low                            &   High              \\ 
      \textbf{Computation Pattern}    & Dynamic \& Irregular           &   Static \& Regular    \\   
      \textbf{Computation Intensity}  & Low                            &   High \\
      \textbf{Execution Bound}        & Memory                         &   Compute \\ 
    \bottomrule
    \end{tabular}}
\end{table}

According to above analysis, hybrid execution patterns exist in GCNs, which are summarized in Table~\ref{tab:variable_pattern}. The \emph{Aggregation} phase performs dynamic and irregular execution pattern, bounded by memory, while the \emph{Combination} phase is static and regular, bounded by computation.

\noindent\textbf{Differences from Conventional Workloads.} 
Beside hybrid execution patterns in GCNs, there are additional characteristics that distinguishes GCNs from conventional workloads. Specifically, in the \textit{Aggregation} phase, the length of feature vectors is variable rather than fixed as in traditional graph analytics, which is determined by the input dataset and MLP structure. Moreover, the length of the feature vectors in each vertex is usually orders of magnitude longer than that of traditional graph analytics. This introduces high intra-vertex parallelism.
In the \textit{Combination} phase, the MLP parameters are fully shared by all vertices while non-reusable in traditional MLP models if not using the batching technique. This induces numerous highly reusable inter-vertex data.
Besides, these two phases are executed alternatively to produce the final result, while conventional workloads iteratively perform only the graph traversal or the neural network propagation.

\subsection{The Need for a GCN Accelerator}
GCNs are showing great potential in various tasks \cite{GraphSage,protein_interface_prediction,web_scale_GCN,GNN_Review,GraphRepresentationLeaning,comprehensive_gnn_survey}. Many companies, such as Google~\cite{Graph_Nets_library}, Facebook~\cite{pytorch_biggraph}, and Alibaba~\cite{AliGraph} have deployed GCNs in data centers, which reflects the increasing importance and scope of upcoming applications. An efficient architecture is timely to achieve high performance and stimulate GCN development.
Therefore, given the above characterizations, we explain our motivation of designing a GCN accelerator.

\noindent\textbf{Design Requirements.} 
Given the characteristics of GCNs, we present the design requirements to perform GCNs with high performance and energy efficiency.
First, \textit{Aggregation} phase demands efforts to alleviate the irregularity that degrades performance. 
On the other hand, \textit{Combination} phase needs more attention to leverage the regularity to improve the intensive computations with better parallelism and faster synchronization.
Second, the high-degree intra-vertex parallelism and the highly reusable inter-vertex data need to be exploited.
Third, to achieve higher performance and energy efficiency, the execution of \textit{Aggregation} phase and \textit{Combination} phase need to be efficiently fused. Unfortunately, existing architectures fail to address these requirements, resulting in the following inefficiencies.

\noindent\textbf{Inefficiencies of General-Purpose Processors.}  
On CPUs, the irregularity in \textit{Aggregation} phase makes GCNs ill-suited to current cache hierarchy design and data prefetching techniques. 
Besides, it is hard to efficiently reuse the highly reusable parameter data between compute units \cite{eyeriss}. 

GPUs are inherently optimized for compute-intensive workloads with regular execution pattern \cite{Tesla} such as neural networks, but handling the \textit{Aggregation} phase with irregular memory accesses suffers from low efficiency. 
Besides, the processing of \textit{Combination} with strong parameter sharing needs costly data copy and thread synchronization. 

Both CPUs and GPUs lack inter-phase optimization for GCN execution.
To leverage the advantages of hardware-optimized functions~\cite{cuBLAS,Pytorch_Scatter}, current programming framework for GCNs usually adopts coarse-grained execution, which results in phase-by-phase execution. This compromises the design space with phase interaction, hindering the improvement beyond the individual optimization for each phase.

\noindent\textbf{Inefficiencies of Conventional Accelerators.}
Specialized accelerators tailored to graph analytics or neural networks gain significant speedup and energy savings compared to general-purpose processors. Whereas, they are inefficient in processing GCNs due to following reasons: i) they are usually only designed to either alleviate irregularity or exploit regularity, while GCNs need both; 
ii) they fail to leverage the new kinds of parallelism and data reuse to further improve performance; iii) single-paradigm design make them hard to fuse the execution of the two phases.

\noindent\textbf{Opportunities for Customization.} Designing a specialized accelerator for a specific domain is an efficient and prevalent solution to address the inefficiencies of existing architectures, since it can tailor the memory hierarchy and computation unit to the specific workload.
For GCNs, we can build an accelerator with a hybrid architecture using different optimizations for the two phases. 
For the \emph{Aggregation} phase, it is possible to obtain the knowledge of graph data in advance and schedule the accesses to alleviate the irregularity. 
Moreover, the computation for each vertex can also be scheduled to exploit edge parallelism and intra-vertex parallelism. 
For the \emph{Combination} phase, we draw inspirations from current neural network accelerators to efficiently perform MVMs in parallel with parameter sharing. Beyond the individual optimizations of the two phases, the serial inter-phase dataflow can be pipelined in finer grain. Moreover, all off-chip memory accesses can be controlled to improve the overall memory access efficiency. Putting all these together, there are huge opportunities to design an efficient GCN accelerator with high performance.

%% file: tex/design.tex
\section{Architecture Design}
In this section, we design \emph{HyGCN} to support the efficient execution of GCNs. We first introduce the programming model and then present details of the architecture design.

\subsection{Edge- and MVM-Centric PM}

The goal of building a programming model (PM) is to exploit available parallelisms and achieve hardware transparency for programmers \cite{AliGraph}. For \emph{Aggregation}, there are gather- and scatter-based processing methods. Since the scatter-based method usually produces large amount of atomic operations and requires a synchronization after the processing of all vertices, the degree of parallelism will be degraded~\cite{Graphicionado}. On the contrary, the gather-based method can control the program behavior easily and preserve the execution parallelism. Therefore, we select the gather-based processing in our design. Nevertheless, this processing mode leads to intensive memory access and vertex computation. To address this problem, we employ an edge-centric PM to exploit the edge-level parallelism. Each vertex possesses many incoming edges (neighbors), which can be aggregated in an edge-by-edge pipeline. In this way, workload for each vertex can be divided into subworkloads and assigned to each computation unit for processing in parallel. 
For \emph{Combination}, the situation is relatively easier. Since the computation of each vertex acts like the MLP, we directly focus on the MVM operations.

Our edge- and MVM-centric PM for GCNs is shown in Algorithm \ref{alg:programming_model}. At each vertex $v \in V$, the sampled neighbor indices are read first, which is a subset of all neighbors. Each index corresponds to an edge connecting $v$ and a neighbor vertex $u$, i.e. $e(u,~v)$. By traversing all sampled edges connected $v$, all the feature vectors of corresponding neighbors can be aggregated onto the feature vector of $v$. Then, a \textbf{\emph{Combine}} function can start performing the \emph{Combination} phase that is comprised of a series of MVMs. In this PM, the edge-level and MVM-level parallelism can be exploited. 

Note that in Algorithm \ref{alg:programming_model} we do not express the Pool and Readout operations explicitly since they are not always needed. In fact, the Pool operation can be represented by two GCNs and additional matrix operations. The GCNs can be performed entirely by the two engines, the matrix transposes can be executed by the flexible \emph{Aggregation} engine, and the matrix multiplications can be executed by the \emph{Combination} engine. The Readout operation can be expressed by an additional single vertex that connects all vertices in the graph, which can be accomplished by the \emph{Aggregation} engine.

\begin{algorithm}[!hptb]
 \label{alg:programming_model}
 \caption{\textbf{\normalsize\emph{Edge- and MVM-centric Programming Model for Aggregation and Combination Phase}}}
 $initial~$\emph{\textbf{SampleNum}}; \\  
 $initial~$\emph{\textbf{SampleIndexArray}}; \\    
 \For{$each~node~v~\in~V$}{
    $agg\_res \leftarrow init()$; \\
    \centerline{{\underline{\color{blue} $\triangleleft \quad \textbf{\emph{Edge-centric Parallelism}}$}}}
    $sample\_idxs \leftarrow \textbf{\emph{SampleIndexArray}}[v.nid]$;\\
    \For{$each~sample\_idx~in~sample\_idxs$}{ 
        $e(u,v) \leftarrow$ \emph{\textbf{EdgeArray}}$[sample\_idx]$;\\
        $agg\_res \leftarrow \textbf{\emph{Aggregate}}(agg\_res, u.feature)$;\\
    }
    \centerline{{\underline{\color{blue} $\triangleleft \quad \textbf{\emph{MVM-centric Parallelism}}$}}}
    $v.feature \leftarrow \textbf{\emph{Combine}}(agg\_res, weights, biases)$;\\
    }
\end{algorithm}

\subsection{Architecture Overview}
\begin{figure}[!hptb] 
    \vspace{-5pt}
    \centering
    \includegraphics[page=2, width=0.48\textwidth]{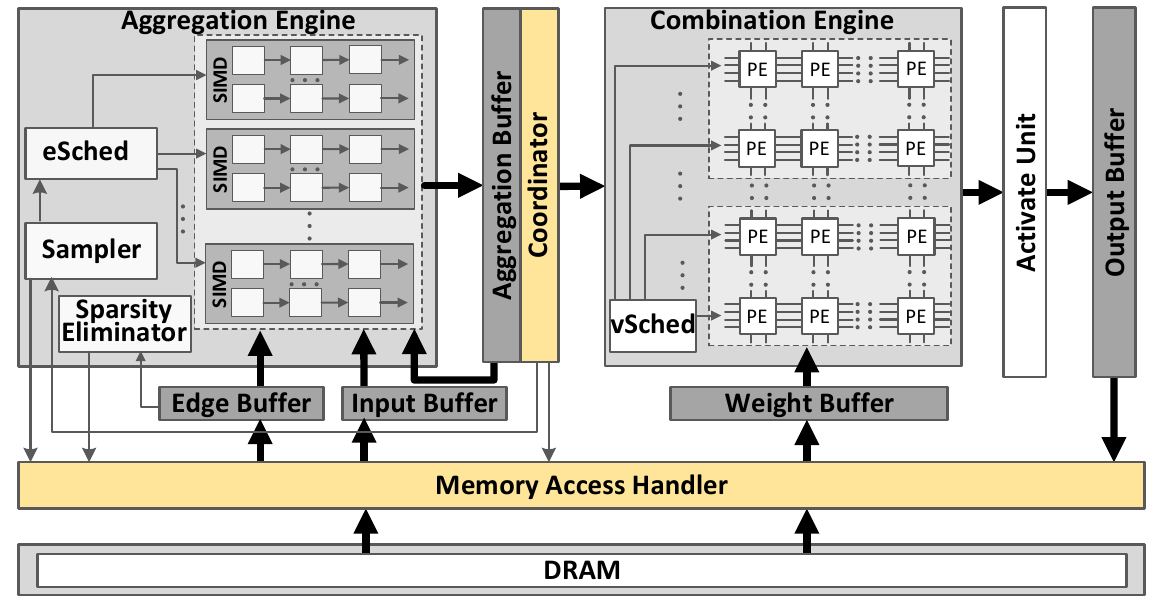}
    \vspace{-25pt}
    \caption{Architecture overview of \emph{HyGCN}.}
    \label{fig:arch}
    \vspace{-8pt}
\end{figure}

Based on the proposed PM, Fig. \ref{fig:arch} depicts the architecture of \emph{HyGCN}. We construct the system using a hybrid architecture, which includes two engines (\emph{Aggregation Engine} and \emph{Combination Engine}) and one memory access handler. A communication interface (\emph{Coordinator}) is introduced to bridge these two engines. Therefore, the interference between them is mitigated and their execution pipeline is established.

The \emph{Aggregation Engine} aims to realize the efficient execution of irregular accesses and computations. To exploit the edge-level parallelism, a task scheduler (\emph{eSched}) is designed to assign the edge processing workloads onto SIMD cores. To support the Sampling operation, we introduce a \emph{Sampler} into the \emph{Aggregation Engine}. The \emph{Sampler} selects edges from the edge list of each vertex using a uniform or predefined distribution in terms of index interval. The former indices for edge sampling are based on dynamic generation while the latter ones are predefined and can be read from off-chip memory like in \cite{FastGCN, AS-GCN}. To reduce the latency of data access, we employ embedded DRAM (eDRAM) to cache various data to improve data reuse. An \emph{Edge Buffer} is used to cache edges to exploit spatial locality in the edge array. An \emph{Input Buffer} is used to cache the vertex features in $X^{k-1}$ and an \emph{Aggregation Buffer} is used to cache the intermediate aggregation results, to exploit temporal locality. To hide the DRAM access latency, both the \emph{Edge Buffer} and \emph{Input Buffer} adopt the double buffer technique. Specifically, we design a \emph{Sparsity Eliminator} to avoid redundant feature loads of the vertices that share no edges with the aggregating vertex.

The \emph{Combination Engine} is designed to maximize the efficiency of regular accesses and computations. In order to improve the processing parallelism and data reuse, we adopt the well-known systolic array design \cite{TPU} and modify it to be compatible with GCNs. A \emph{Weight Buffer} is used to cache the weight matrix to exploit their temporal locality, and an \emph{Output Buffer} is used to coalesce the write accesses of the final features. Similarly, they also leverage the double buffer technique to hide off-chip access latency. The \emph{Combination} engine takes the aggregation result of each vertex $v$ from the \emph{Aggregation} engine and the weight matrix from the \emph{Weight Buffer} as inputs to execute the MVM operation. The \emph{vSched} is responsible for the workload assignment. After the MVM operations, an activation operation is performed by \emph{Activate Unit} to produce the new feature vector of vertex $v$. Different from normal systolic array, our systolic array is multi-granular that can be used as multiple smaller arrays or a whole large array under different optimization scenarios.

To improve the bandwidth utilization, a prefetcher is designed to explicitly prefetch graph data and parameter data. For example, the prefetching of the feature vectors is as follows.
The prefetcher first prefetches the edges of current processing vertices. 
After receiving these edges, \textit{Sparsity Eliminator} obtains the indices of neighbors from these edges and sends them to the prefetcher. The prefetcher uses them to prefetch the feature vectors immediately.

\subsection{Aggregation Engine} \label{sec:aggregation_engine}

To optimize the computation of \emph{Aggregation}, we introduce a vertex-disperse processing mode.
To optimize memory accesses, we employ a static graph partition method to enhance data reuse and a dynamic sparsity elimination technique to reduce unnecessary data accesses.

\subsubsection{Execution Mode}

\begin{figure}[!hptb] 
    \vspace{-8pt}
    \centering
    \includegraphics[page=6, width=0.48\textwidth]{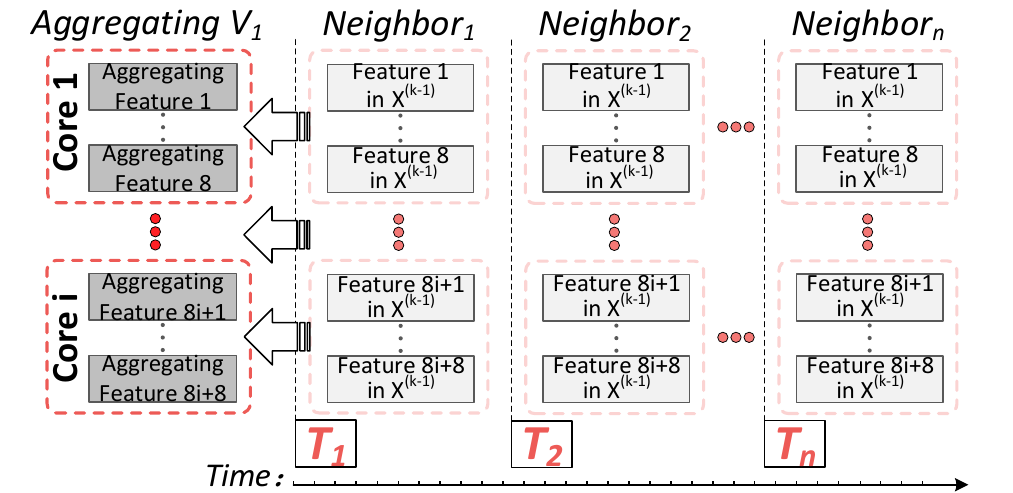}
    \vspace{-20pt}
    \caption{Vertex-disperse processing mode where the workloads of each vertex are assigned to all SIMD cores.}
    \label{fig:aggregation_flow}
    \vspace{-10pt}
\end{figure}

There are two processing modes for SIMD cores to process edges in parallel. The first one is vertex-concentrated, where the workloads of each vertex are assigned to a single SIMD core. This mode can produce the aggregated features of vertices in burst mode, i.e. periodically processing a group of vertices. However, the processing latency of a single vertex (termed as vertex latency) is long, and the fast vertices have to wait for the slow vertices leading to workload imbalance. 
Furthermore, it also loses the parallelism that the aggregation of each element can be performed in parallel (i.e., intra-vertex parallelism).
Therefore, we use the second processing mode, which is shown in Fig. \ref{fig:aggregation_flow}. It assigns the aggregation of elements inside the vertex feature vector of each vertex to all cores, termed as vertex-disperse mode.
If a vertex cannot occupy all cores, free cores can be assigned to other vertices. Thus, all cores are always busy without workload imbalance.
Moreover, since the intra-vertex parallelism has been exploited, the vertex latency for a single vertex is smaller than processing multiple vertices together. Furthermore, it also enables the immediate processing of each vertex in the following \emph{Combination Engine}.

\begin{figure*}[!hptb] 
    \centering
    \includegraphics[page=4, width=\linewidth]{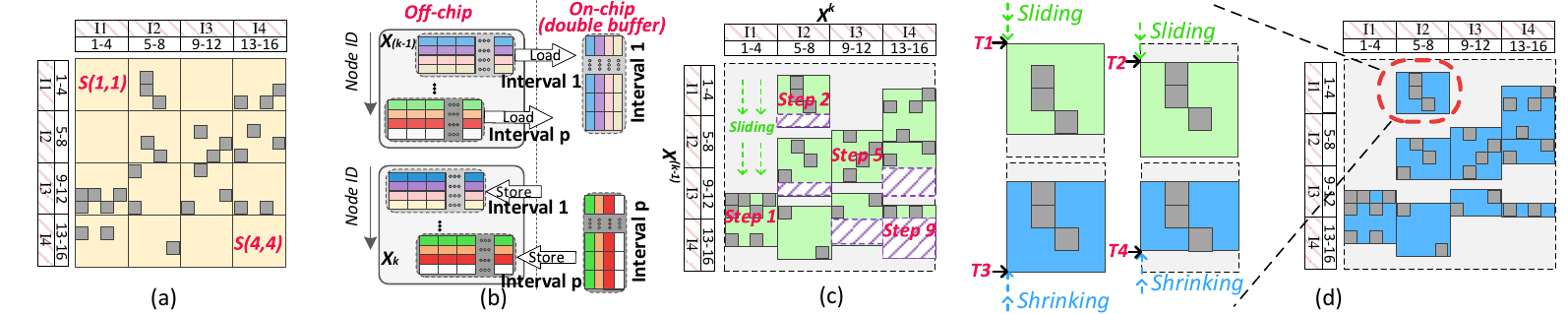}
    \vspace{-20pt}
    \caption{Static graph partition for data reuse and dynamic sparsity elimination to reduce redundant accesses: (a) interval-shard partition; (b) interval-wise feature access; (c) window sliding; (d) window shrinking.}
    \label{fig:data_partition}
     \vspace{-10pt}
\end{figure*}

\subsubsection{Graph Partitioning (Static)}

We borrow the abstraction of vertex interval and edge shard from \cite{graphchi,NXgraph} to partition graph data, which is the basis of our data-aware sparsity elimination in the next subsection. We do not need explicit preprocessing to generate the intervals and shards since we directly take the data format of compressed sparse column (CSC) as input. As exampled in Fig. \ref{fig:data_partition}(a), the 16 vertices are organized as several intervals (i.e. from $I_1$ to $I_4$, each with four vertices), and the edges are organized as 4$\times$4 shards (i.e. from $S(1,1)$ to $S(4,1)$, each with 16 edges at most). The intervals and shards are disjoint.

The feature vector length of each vertex is usually large, so exploiting the locality of features is critical. We group the vertices within the same interval together (e.g. $I_i$) and then process the aggregation of their source neighbors also interval by interval (i.e. traverse $I_j$), as expressed in Algorithm \ref{alg:pm_interval}. Based on this flow, the feature accesses of all vertices in an interval are merged (see Fig. \ref{fig:data_partition}(b)). The resulting benefits are twofold. First, the vertices in $I_i$ usually have overlapped neighbors in $I_j$, therefore, the loaded feature data of $I_j$ can be reused when performing feature aggregation. Second, when traversing all $I_j$, the intermediate aggregated results of $I_i$ are remained in buffer which can also be reused when performing feature update. In practice, edge shards usually are not square as our simplified illustration in Fig. \ref{fig:data_partition}. The shard height is determined by the capacity of \emph{Input Buffer}, while the shard width is determined by the capacity of \emph{Aggregation Buffer}. The \emph{Edge Buffer} size affects both height and width since it accommodates all edges of each shard.

\subsubsection{Data-Aware Sparsity Elimination (Dynamic)}

With the data reuse optimization, we further attempt to reduce the redundant accesses since the graph connections are sparsely distributed. To eliminate the sparsity, we propose a window-based sliding and shrinking approach. The key idea is that we first slide the window (with the same size of an edge shard) downward until an edge appears in the top row, and then we shrink the window size by moving the bottom row upward until an edge is met.

\noindent\textbf{Window Sliding.} Fig. \ref{fig:data_partition}(c) illustrates the window sliding process. For each vertex interval, the top shard window gradually slides downward. It will not stop until an edge appears on its top row. Then a new window with the same size is created, whose top row follows the bottom row of its previous window. The stop criterion is the same for every window. In this way, windows continuously arise, slide downward, and stop. All the positions where windows stop are recorded as effectual shards.

\vspace{-5pt}
\begin{algorithm}[!hptb]
 \SetKwRepeat{Do}{do}{while}
 \label{alg:pm_interval}
 \caption{\textbf{\normalsize\emph{Interval-wise Aggregation}}}

    \For{$each~interval~I_{i}~in~X^{k}$}{
    $agg\_res \leftarrow init()$;\\
    \For{$each~interval~I_{j}~in~X^{(k-1)}$}{ 
        $agg\_res \leftarrow$ \emph{\textbf{Aggregation}}$(I_{j}, agg\_res)$;\\
    }
    $I_{i} \leftarrow$ \emph{\textbf{Combination}}$(agg\_res)$;\\
    }
\end{algorithm}
\vspace{-10pt}

\noindent\textbf{Window Shrinking.} Although the window sliding can capture most effectual edges, sparsity still exists on the bottom side (within the purple dashed boxes). This is because the above sliding direction is downward. To reduce this part of sparsity, we propose window shrinking here. Specifically, the bottom row of each recorded window moves upward until it meets an edge, and then the window shrinks. Fig. \ref{fig:data_partition}(d) illustrates the sliding and shrinking process of one window in detail and gives the final recorded effectual shards. Different from previous partition, the sizes of final shards are usually different due to the window shrinking. 

\vspace{-5pt}
\begin{algorithm}[!hptb]
 \SetKwRepeat{Do}{do}{while}
 \label{alg:pm_interval_sparsity}
 \caption{\textbf{\normalsize\emph{Interval-wise Aggregation with Sparsity Elimination}}}

    \For{$each~interval~I_{i}~in~X^{k}$}{
    $row\_pos \leftarrow 1$;\\
    $agg\_res \leftarrow init()$;\\
    \Do{$(I_{j}$ \emph{!=} $\varnothing)$}{ 
        ($I_{j},row\_pos) \leftarrow$ \emph{\textbf{GetOneEffectInterval}}$(~X^{(k-1)}, A, I_{i}, row\_pos)$;\\
        $agg\_res \leftarrow$ \emph{\textbf{Aggregation}}$(I_{j}, agg\_res)$;\\
     }
    
    $I_{i} \leftarrow$ \emph{\textbf{Combination}}$(agg\_res);$\\
    }  
\end{algorithm}
\vspace{-5pt}

Given the effectual shards after sparsity elimination, the execution flow of \emph{Aggregation} follows Algorithm \ref{alg:pm_interval_sparsity}. The only difference from Algorithm \ref{alg:pm_interval} is that the each neighbor interval $I_j$ is dynamically determined by window sliding and shrinking (see Algorithm \ref{alg:sliding_shrinking}). The starting row of each neighbor interval varies due to sliding and the interval length in the row dimension also varies due to shrinking. In this way, only the feature data of remaining neighbor vertices when performing the aggregation operation for each interval $I_i$ are loaded, which eliminates plenty of redundant accesses.

\begin{algorithm}[!hptb]
 \SetKwRepeat{Do}{do}{while}
 \label{alg:sliding_shrinking}
 \caption{\textbf{\normalsize\emph{GetOneEffectInterval}}}
    \centerline{{\underline{\color{blue} $\triangleleft \quad \textbf{\emph{Window Sliding}}$}}}
    \While{$(edge(row\_pos,v) == \varnothing~for~\forall v \in I_{i})$}
    {
     $row\_pos \leftarrow row\_pos + 1$;\\ 
    }
    $win_{start} \leftarrow row\_pos$;\\ 
    $win_{end} \leftarrow row\_pos + Window_{height} - 1$;\\ 
    $row\_pos \leftarrow win_{end} + 1$;\\ 
    \centerline{{\underline{\color{blue} $\triangleleft \quad \textbf{\emph{Window Shrinking}}$}}}
    \While{$(edge(win_{end},v) == \varnothing~for~\forall v \in I_{i})$}{
     $win_{end} \leftarrow win_{end} - 1$;\\ 
    }
    $I_{effectual} \leftarrow X^{(k-1)}[win_{start}:win_{end}]$;\\
    $\textbf{return}~I_{effectual}$;\\
\end{algorithm}
\vspace{-5pt}

Compared to traditional graph analytics, the feature data reuse from graph partitioning and redundant access reduction from sparsity elimination in GCNs are considerable efforts. This is because the feature of each vertex in GCNs is a vector with thousands of elements, while the feature data in traditional graph analytics are small, usually with one element for each vertex.
Besides, our optimization achieves more when the Sampling operation is used, which increases sparsity since only sampled neighbors are required during \emph{Aggregation}.

\subsection{Combination Engine}

\begin{figure}[!hptb] 
    \centering
    \includegraphics[page=7, width=0.48\textwidth]{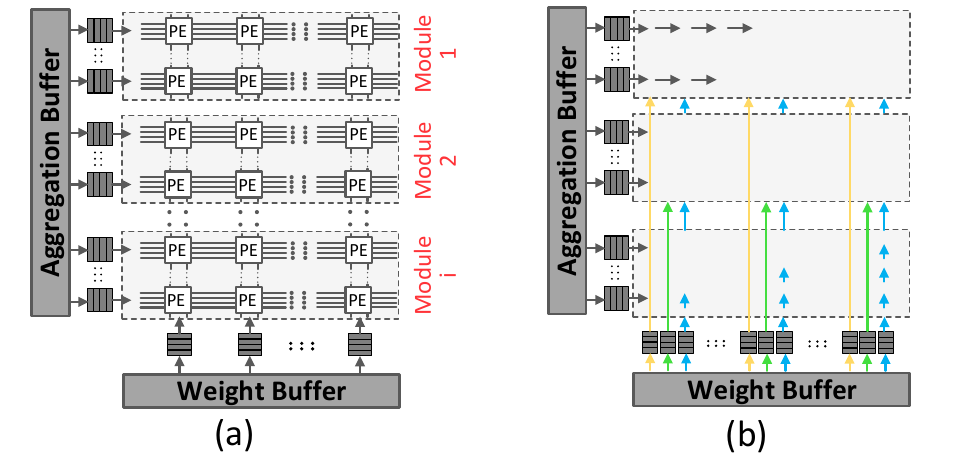}
    \vspace{-25pt}
    \caption{\emph{Combination Engine} design: (a) multiple systolic modules; (b) different dataflow patterns. }
    \label{fig:hierarchy_systolic_array}
    \vspace{-13pt}
\end{figure}

The \emph{Combination} operation at each vertex acts like a neural network, the execution of which is regular but compute-intensive. Our design is based on the well-known systolic array. To adapt it for the two processing modes of \emph{Aggregation Engine} (see Fig. \ref{fig:aggregation_flow}), we integrate multiple arrays rather than a single one, as shown in Fig. \ref{fig:hierarchy_systolic_array}(a). A group of systolic arrays is assembled to form a systolic module. We allow a multigranular use of these systolic modules, including the independent working mode and cooperative working mode.

\subsubsection{Independent Working Mode}

\begin{figure}[!hptb]
    \centering
    \includegraphics[page=8, width=0.46\textwidth]{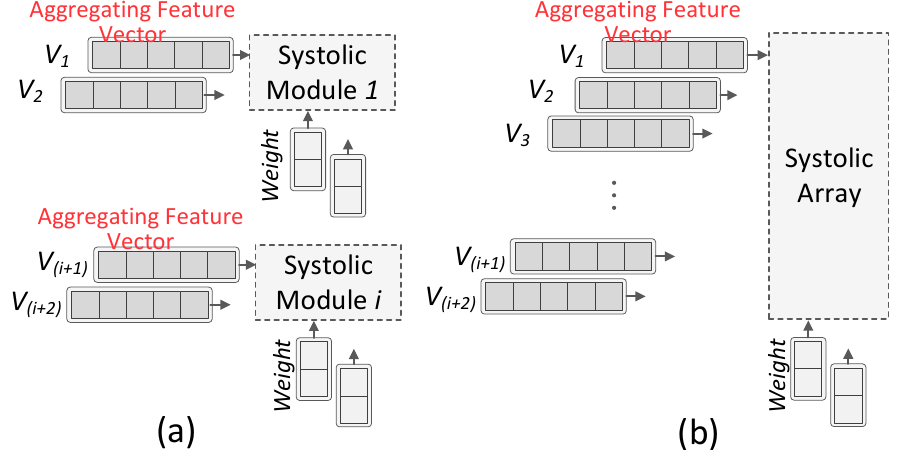}
    \vspace{-10pt}
    \caption{Different use of the systolic arrays: (a) independent working mode; (b) cooperative working mode.}
    \label{fig:work_mode_systolic_array}
    \vspace{-10pt}
\end{figure}

In this mode, the systolic modules work independently from each other. Each of them processes the MVM operations of a small group of vertices, as illustrated in Fig. \ref{fig:work_mode_systolic_array}(a). The weight parameters for each module in this case are directly accessed from the \emph{Weight Buffer} and just reused within module, as depicted in Fig. \ref{fig:hierarchy_systolic_array}(b). The advantage of this mode is the lower vertex latency because we can process the \emph{Combination} operations of this small group of vertices immediately once their aggregated features are ready, without waiting for more vertices. This mode matches well with the vertex-disperse processing mode of \emph{Aggregation Engine} in Fig. \ref{fig:aggregation_flow}, where the aggregated features are produced quickly but sequentially.

\subsubsection{Cooperative Working Mode}

Besides working separately, these systolic modules can be further assembled together to simultaneously process more vertices, as shown in Fig. \ref{fig:work_mode_systolic_array}(b). Different from the immediate processing of vertices, this mode requires to assemble the aggregated features of a large group of vertices together before performing their \emph{Combination} operations. The advantage is that, the weight parameters can flow from the \emph{Weight Buffer} to the downstream systolic modules and then gradually to the upstream ones (see Fig. \ref{fig:hierarchy_systolic_array}(b)), which are greatly reused by all systolic arrays. This helps reduce the energy consumption.

No matter which working mode is selected in the \emph{Combination Engine}, the weights can be reused inherently in \emph{Weight Buffer} when processing different vertices. However, in traditional neural networks, especially MLPs, the weights cannot be shared without batching technique. The multi-granular systolic array design is also specific to our architecture in order to accommodate different application needs.

\subsection{Inter-Engine Optimization}

To efficiently fuse the phase-by-phase execution, we orchestrate the execution pipeline and DRAM access of \emph{Aggregation} engine and \emph{Combination} engine by the \emph{Coordinator}.

\subsubsection{Latency- or Energy-Aware Pipeline}

\noindent\textbf{Ping-Pong Aggregation Buffer.} To reuse the aggregation results produced by the \emph{Aggregation} engine, we add an \emph{Aggregation Buffer} between the two engines. This buffer can be written by the \emph{Aggregation Engine} and can be read by the \emph{Combination Engine}. Before the final aggregated results are generated, the \emph{Aggregation Buffer} stores the partial results that will be read by the \emph{Aggregation Engine} for feature accumulation. In order to increase the parallelism of these two engines, we implement a ping-pong buffering mechanism where the \emph{Aggregation Buffer} is split into two chunks. In this way, the executions of aggregation and combination are decoupled, which enables an inter-engine pipeline. 

To accommodate the needs of different applications, we provide two pipeline modes as follows. \par
\begin{figure}[!hptb] 
    \centering
    \includegraphics[page=9, width=0.46\textwidth]{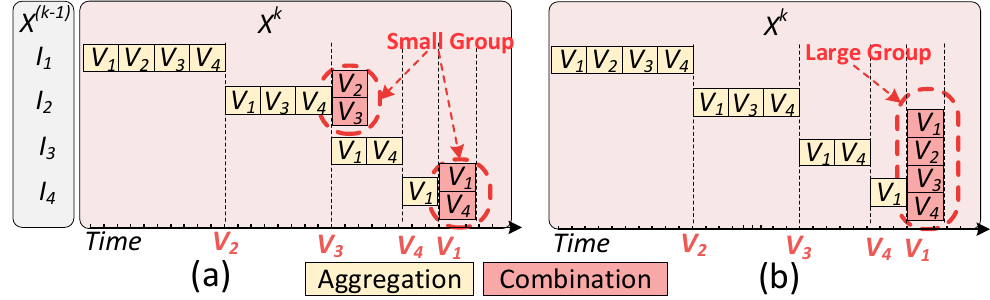}
    \vspace{-10pt}
    \caption{Timing illustration of different pipeline modes: (a) latency-aware pipeline; (b) energy-aware pipeline.}
    \label{fig:pipeline_mode}
    \vspace{-10pt}
\end{figure}

\noindent\textbf{Latency-Aware Pipeline.} In this pipeline mode, the \emph{Combination Engine} works in the systolic module independent mode. The aggregated features are produced vertex by vertex in the \emph{Aggregation Engine}, and the following combination will be processed immediately once the aggregated features of a small group of vertices are ready. Therefore, the average processing latency for each vertex can be lower. The overall timing is illustrated in Fig. \ref{fig:pipeline_mode}(a), where $V$ denotes the vertices for aggregation, and $I$ represents the neighbor intervals.

\noindent\textbf{Energy-Aware Pipeline.} 
The energy-aware pipeline uses the systolic module cooperative mode in the \emph{Combination Engine}. The vertex-by-vertex processing changes to a burst mode, where a large group of vertices will be processed together every time. Although the vertex latency is longer, the energy consumption can be reduced due to the weight propagation in the merged systolic arrays without redundant accesses. Fig. \ref{fig:pipeline_mode}(b) presents its timing sequence.

\subsubsection{Coordination of Off-chip Memory Access}

It is hard to determine the memory bandwidth ratio between the two engines since the practical workloads usually vary between \emph{Aggregation} and \emph{Combination}. 
Moreover, the separation of memory systems will increase the configuration overheads and cause bandwidth waste. This is the reason why we use only one off-chip memory. 

\begin{figure}[!hptb] 
    \centering
    \includegraphics[page=10, width=0.46\textwidth]{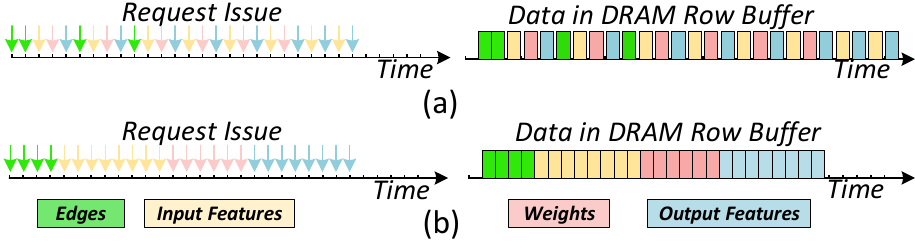}
    \vspace{-10pt}
    \caption{Coordination of off-chip memory access.}
    \label{fig:off_chip_memory_coordination}
    \vspace{-5pt}
\end{figure}

Both the two engines access this memory at runtime, which causes a frequent switching of access locations, leading to inefficiencies.
In total, there are four buffers (\emph{Edge Buffer} \& \emph{Input Buffer} in \emph{Aggregation Engine}, and \emph{Weight Buffer} \& \emph{Output Buffer} in \emph{Combination Engine}) that will be used for accessing the off-chip memory. Due to the interval processing and pipeline mechanism, these accesses usually come concurrently as shown in Fig. \ref{fig:off_chip_memory_coordination}(a). 
If we sequentially handle these access requests, the discontinuous addresses greatly degrade the utilization of row buffer within DRAM.

To solve this problem, we predefine an access priority ($edges>input features>weights>output features$) to assemble the discontinuous requests shown in Fig. \ref{fig:off_chip_memory_coordination}(b). The motivation in using this priority is based on the access sequence when processing a vertex. The access requests are executed batch-by-batch. Therefore, low-priority accesses in the current batch are handled before high-priority accesses coming at the next batch, rather than always high-priority accesses first.
With the improved continuity, the utilization of row buffer can be significantly enhanced. 
Next, we remap these reordered addresses to index the channel and bank using low bits. 
In this way, the memory channel- and bank-level parallelism can be further exploited.

%% file: tex/methodology.tex
\section{Evaluation Results}

We first describe our experimental setup in Section \ref{sec:setup}. Next, to demonstrate the advantages of our design, we compare \emph{HyGCN} to the state-of-the-art software framework in Section \ref{overall_result}. Next, we give the detailed analysis of our optimization techniques in Section \ref{optimization_analysis}. Finally, we present a scalability exploration of our architecture in Section \ref{scalability_exploration}.

\subsection{Experimental Setup}\label{sec:setup}

\noindent\textbf{Methodology}. The performance and energy of \emph{HyGCN} are measured by using the following tools.

\emph{Architecture Simulator.} 
We design and implement a cycle-accurate and execution-driven simulator to measure execution time in number of cycles. This simulator models the microarchitectural behaviors of each module, which is integrated with Ramulator \cite{Ramulator} to simulate the behaviors of memory accesses to High Bandwidth Memory (HBM). 

\emph{CAD Tools.} For the measurements of area, power, and critical path delay (in cycles) for each module, we implement and synthesize each module in Verilog. We use the Synopsys Design Compiler with the TSMC 12 nm standard VT library for the synthesis, and estimate the power 
using Synopsys PrimeTime PX. The slowest module has a critical path delay of 0.9 ns including the setup and hold time, putting the \emph{HyGCN} 
comfortably at 1 GHz clock frequency. 

\emph{Memory Measurements.} The area, power, and access latency of the on-chip scratchpad memory are estimated using Cacti 6.5 \cite{CACTI}. Since Cacti only supports down to 32 nm technologies, we apply four different scaling factors to convert them to 12 nm technology as shown in \cite{technology_scale,ozdal_energy_2016}.  The energy of HBM 1.0 is estimated with 7 pJ/bit as in \cite{HBM_Power,GraphDynS}.

\begin{table}[!htpb]
	\caption{Dataset information \cite{KKMMN2016, dataset_explanation}.}\label{table:dataset}
	\centering
	\renewcommand\arraystretch{1.2}
    \resizebox{0.46\textwidth}{!}{
		\begin{tabular}{*6{c}}
	     	\toprule   
			\textbf{Dataset}          & \textbf{\#Vertex}  &  \textbf{Feature Length}  & \textbf{\#Edge}    & \textbf{Storage}  \\   
			 IMDB-BIN (IB)   & 2,647       & 136             & 28,624      &  1.5MB \\ 
			 Cora (CR)       & 2,708       & 1,433           & 10,556      &  15MB  \\  
			 Citeseer (CS)   & 3,327       & 3,703     	     & 9,104       &  47MB  \\  
  			 COLLAB (CL)     & 12,087      & 492             & 1,446,010   &  28MB  \\ 
  			 Pubmed (PB)     & 19,717      & 500             & 88,648      &  38MB  \\ 
  			 Reddit (RD)      & 232,965     & 602             & 114,615,892 &  972MB          \\  
             \bottomrule
		\end{tabular}
	}
\end{table}

\begin{table}[!htbp]
    \vspace{-15pt}
	\caption{Configuration of convolution layers. Here |$a^k_v$| denotes the length of feature vector $a^k_v$. }\label{table:gcn_model}
	\centering
    \renewcommand\arraystretch{1.3}
    \resizebox{0.48\textwidth}{!}{
    \begin{tabular}{ccc}
    \toprule
                      & \textbf{\#Sampling Neighbors}    &  \textbf{Aggregation \& Combination (MLP)}                              \\ \midrule
    \textbf{GCN (GCN)}          &  ---                   &  Add  \& |$a^k_v$|--128                 \\ 
    \textbf{GraphSage (GSC)}    &  25                    &  Max \& |$a^k_v$|--128                 \\ 
    \textbf{GINConv (GIN)}      &  ---                   &  Add  \& |$a^k_v$|--128--128            \\ \midrule
\multirow{2}{*}{\textbf{DiffPool (DFP)}} 
                      &  \textbf{GCN$_{pool}$}           & \textbf{GCN$_{embedding}$}              \\ \cmidrule{2-3}
                      & Min \& |$a^k_v$|--128           & Min \& |$a^k_v$|--128                  \\
                                                                                                   \bottomrule
\end{tabular}
}
\vspace{-5pt}
\end{table}

\noindent\textbf{Benchmark Graph Datasets and GCN Models.} Table \ref{table:dataset} and Table \ref{table:gcn_model} provide the information of the benchmark graph datasets and GCN models used in our evaluation. The datasets in Table \ref{table:dataset} are standard ones in the GCN domain. They are actually not small although the number of vertices is smaller than that used in conventional graph analytics, due to the long length of feature vectors.
On CPU, the datasets with more than one graphs are tested by assembling randomly selected 128 graphs into a large graph before processing for GCN, GSC, and GIN or batching the same number of graphs for DFP. 
On \emph{HyGCN}, the testing methods remain the same with CPU except that the selected graphs for DFP are processed one by one rather than in a batched mode. 

\noindent\textbf{Baseline Platform.} To compare the performance and energy consumption of \emph{HyGCN} with state-of-the-art works, we evaluate PyTorch Geometric (PyG) \cite{PyTorch_Geometric} on a Linux workstation equipped with two Intel Xeon E5-2680 v3 CPUs and 378 GB DDR4 memory and on an NVIDIA V100 GPU, denoted as PyG-CPU and PyG-GPU, respectively. 
Table \ref{table:acc_system} lists the system configurations for above implementations. 

\begin{table}[!htbp]
    \vspace{-10pt}
	\caption{\textbf{System configurations.}}\label{table:acc_system}
	\centering
    \resizebox{0.48\textwidth}{!}{
    \begin{tabular}{cccc}
    \toprule
    & \textbf{PyG-CPU}  & \textbf{PyG-GPU}  & \textbf{\textit{HyGCN}}            
    \\ \midrule       
    \begin{tabular}[c]{@{}c@{}}\textbf{Compute}\\ \textbf{Unit}\end{tabular}    
    & \begin{tabular}[c]{@{}c@{}}2.5 GHz @\\ 24 cores\end{tabular} & \begin{tabular}[c]{@{}c@{}}1.25Ghz @ \\ 5120 cores\end{tabular}       
    & \begin{tabular}[c]{@{}c@{}}1 GHz @ 32 SIMD16 cores and\\ 8 systolic modules (each with 4$\times$128 arrays)\end{tabular}  
    \\ \midrule  
    \begin{tabular}[c]{@{}c@{}}\textbf{On-chip}\\ \textbf{Memory}\end{tabular}  & 60MB   & 34MB    & \begin{tabular}[c]{@{}c@{}}128 KB (Input), 2 MB (Edge), 2 MB (Weight),\\ 4 MB (Output) and 16 MB (Aggregation)\end{tabular}   \\ \midrule  
    \begin{tabular}[c]{@{}c@{}}\textbf{Off-chip}\\ \textbf{Memory}\end{tabular}  
    & \begin{tabular}[c]{@{}c@{}}136.5GB/s\\ DDR4\end{tabular}     & \begin{tabular}[c]{@{}c@{}}$\sim$900GB/s \\ HBM$\sim$2.0\end{tabular} 
    & \begin{tabular}[c]{@{}c@{}}256GB/s \\ HBM$\sim$1.0\end{tabular}  \\ \bottomrule
    \end{tabular}      
}
\scriptsize
\par Note: GPU's on-chip memory includes the register files, and L1 and L2 caches. 
\end{table}

%% file: tex/results.tex
\begin{figure*}[!htbp] 
    \centering
    \includegraphics[page=1,width=\textwidth]{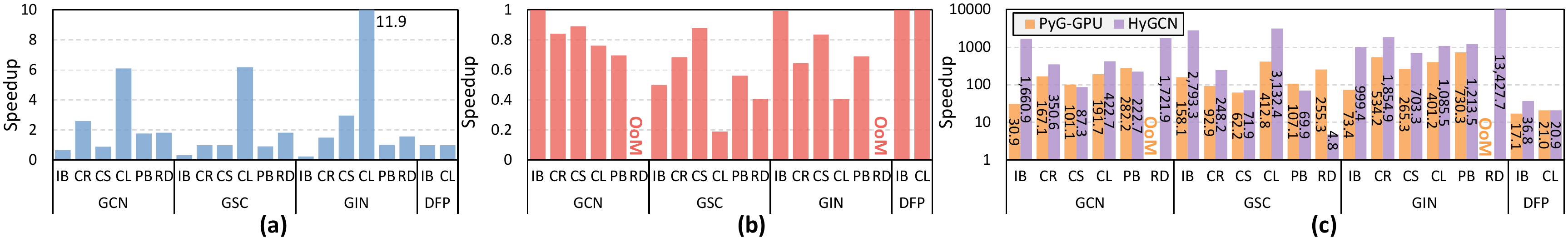}
    \vspace{-23pt}
    \caption{Comparistion to PyG-CPU and PyG-GPU: Speedup of our algorithm optimization on (a) CPU and (b) GPU; (c) Speedup over the optimized PyG-CPU. OoM means the evaluation fails in running on GPU due to out of memory.
    }
    \label{fig:overall_speedup}
    \vspace{-13pt}
\end{figure*}

\subsection{Overall Results}\label{overall_result}
We first apply our algorithm optimization on PyTorch Geometric. And then, we compare our work (\textit{HyGCN}) with PyG-CPU and PyG-GPU in terms of speedup, energy consumption, utilization of DRAM bandwidth, and DRAM access. Finally, the area and power of our design is presented. 

\noindent$\bullet$ \underline{\emph{\textbf{Algorithm Optimization on PyG Framework.}}} 
To show the effect of our algorithm optimization on CPU and GPU platforms, we implement our algorithm optimization proposed in Section~\ref{sec:aggregation_engine} on PyG framework.
The graph is partitioned into multiple shards and they are executed shard by shard (see Fig. \ref{fig:data_partition}(a)). The number of partitions is determined by the capacity of L2 Cache and the length of feature vectors. Note that, PyG leverages the Pytorch Scatter library \cite{Pytorch_Scatter} for the acceleration of \textit{Aggregation} on both CPU and GPU. It helps eliminate the sparsity and exploit the edge parallelism by executing each vertex's \textit{Aggregation} in a hardware thread. Furthermore, the hardware-optimized libraries such as Intel MKL \cite{MKL} and NVIDIA cuBLAS library \cite{cuBLAS} are used to accelerate \textit{Combination} on CPU and GPU, respectively. 

Fig.~\ref{fig:overall_speedup}(a) shows the speedup of PyG-CPU with our algorithm optimization (PyG-CPU-OP) over the naive one without optimization. Thanks to the algorithm improvement, PyG-CPU-OP achieves 2.3$\times$ speedup on average. The performance benefits come from the reduction of frequent replacement of feature vectors since the reusable features after graph partition and the intermediate results of \textit{Aggregation} are buffered in L2 Cache. Fig.~\ref{fig:overall_speedup}(b) presents the same testing on GPU. The performance of PyG-GPU-OP degrades since only a small amount of vertices are processed for each graph partition, which cannot fully utilize thousands of hardware threads on GPU and miss the core advantage of GPU to hide the access latency through many parallel threads. As a result, it is inefficient for GPU to exploit our optimization to improve performance. The optimized PyG-CPU and the naive PyG-GPU are used as baselines in the following evaluation.

\noindent$\bullet$ \underline{\emph{\textbf{Speedup.}}} 
Fig.~\ref{fig:overall_speedup}(c) depicts that \emph{HyGCN} achieves average 1509$\times$ and 6.5$\times$ speedup compared with PyG-CPU and PyG-GPU, respectively. 
The performance improvement comes from the individual optimizations in \emph{Aggregation Engine} \& \emph{Combination Engine}, and the inter-engine pipeline \& coordination. First, the parallel processing in SIMD cores and systolic arrays speed up the computations. Second, the graph partition and sparsity elimination increase the feature reuse and decrease redundant accesses in \emph{Aggregation Engine}, which saves DRAM bandwidth. Third, the weight parameters are reused efficiently in \emph{Combination Engine}, which also helps better utilize the bandwidth. Finally, the inter-engine pipeline further optimizes the parallelism and the off-chip memory access coordination improves the DRAM access efficiency.

For PyG-CPU and PyG-GPU, abundant DRAM accesses and synchronization overheads lead to performance degradation. Specifically, the high randomness of neighbor indices results in poor locality of neighbors' feature vectors, causing many unnecessary DRAM accesses. From the perspective of computation, PyG-CPU and PyG-GPU leverage the hardware-optimized functions (such as scatter~\cite{Pytorch_Scatter} and matrix multiplication~\cite{cuBLAS}) to perform GCNs in a coarse-grained fashion. Although it is the best way to utilize CPU and GPU, it loses the inter-phase parallelism and produces redundant operations. The delay for data copy and synchronization between threads further degrades the performance.

In term of models, GIN achieves better performance than others. The underlying reason is that GIN executes \textit{Aggregation} first on PyG-CPU and PyG-GPU, which introduces abundant computations and accesses since the feature vector size is an order of magnitude larger than that after \textit{Combination}. 
By contrast, other models execute \textit{Combination} first, which greatly reduces the feature length before performing \textit{Aggregation}. This difference causes the inefficient execution of GIN on CPU and GPU, while our \emph{HyGCN} can maintain the performance to a great extent due to the parallel processing and data reuse. For DFP, it includes three matrix multiplications (see Equation (\ref{equation_diffpool})) that can be efficiently executed on CPU and GPU. Therefore, our speedup when performing DFP is relatively lower. The GSC model consumes significant time on the Sampling operation in a preprocessing step, which is not included in the result of PyG-CPU and PyG-GPU. For example on the RD dataset, the preprocessing can cost up to 15 seconds while the execution time is only 0.65 second on PyG-CPU and 0.0025 second on PyG-GPU. In our work, the Sampling operation is executed together with \textit{Aggregation} and considered in the reported result. Thus, the performance of our work is lower than PyG-GPU in Fig. \ref{fig:overall_speedup}(c) but the overall execution time ratio is 0.136 second v.s. 15.7 seconds.

\begin{figure}[!hptb] 
    \centering
    \centering
    \includegraphics[page=1,width=0.43\textwidth]{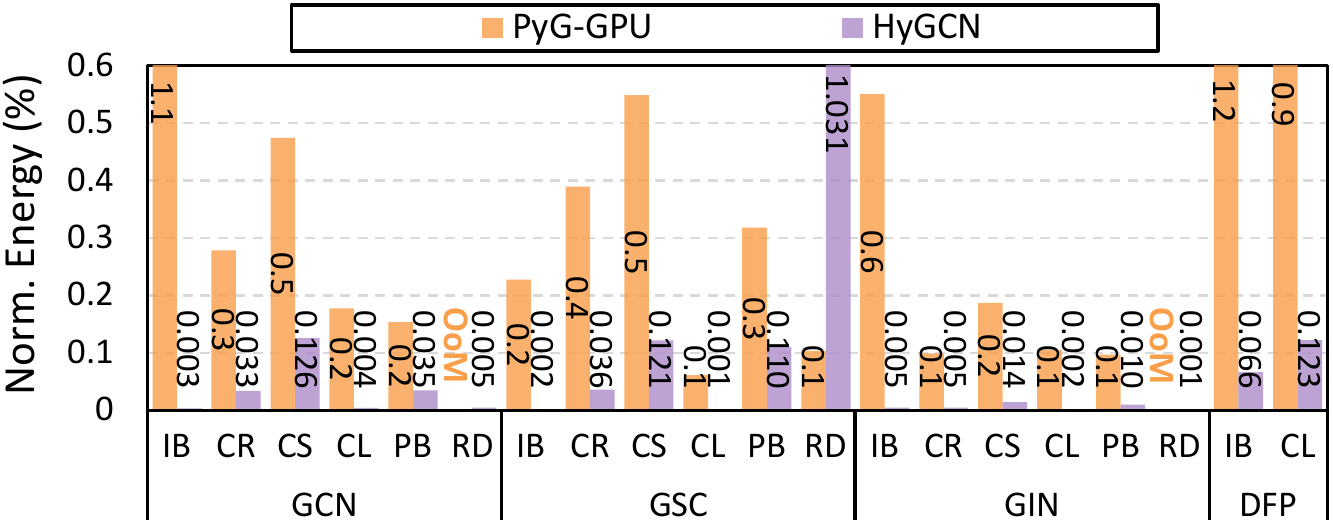}
    \vspace{-12pt}
    \caption{Normalized energy over PyG-CPU.}
    \label{fig:overall_energy}
    \vspace{-10pt}
\end{figure}

\noindent$\bullet$ \underline{\emph{\textbf{Energy Consumption.}}} As Fig.~\ref{fig:overall_energy} shows, \emph{HyGCN} consumes only 0.04\% and 10\% energy on average compared to PyG-CPU and PyG-GPU, respectively. 
The energy consumption of all platforms includes the off-chip memory. Note that, although the results of PyG-CPU and PyG-GPU do not include the overhead of the Sampling operation, they are still costly. For example, the Sampling energy of GSC is 2715J on the RD dataset. In contrast, our work consumes only 1.79J 
compared to the total 2716J in PyG-GPU.

\begin{figure}[!hptb] 
    \vspace{-5pt}
    \centering
    \includegraphics[page=1, width=0.43\textwidth]{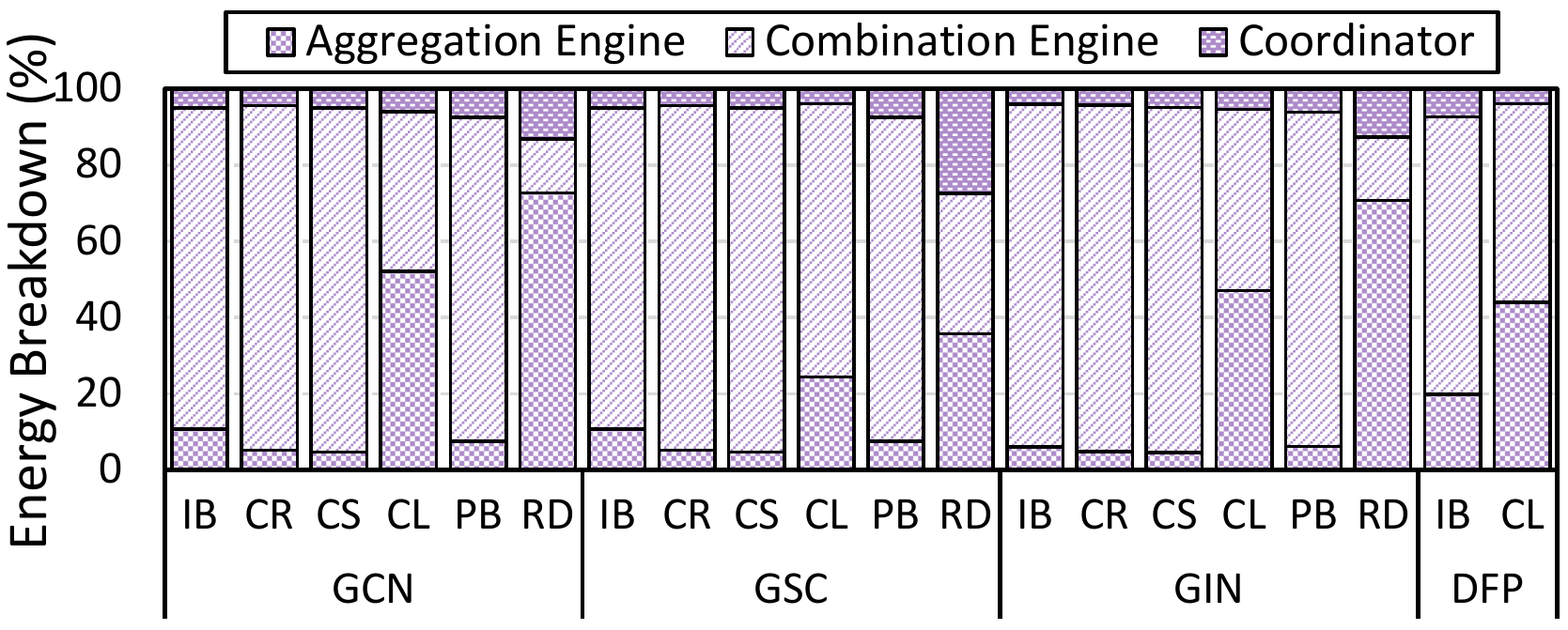}
    \vspace{-15pt}
    \caption{Energy breakdown of \textit{HyGCN}.}
    \label{fig:energy_breakdown}
    \vspace{-5pt}
\end{figure}

\begin{figure*}[!htbp] 
    \centering
    \begin{minipage}{0.56\textwidth}
    \centering
    \includegraphics[page=1,width=\textwidth]{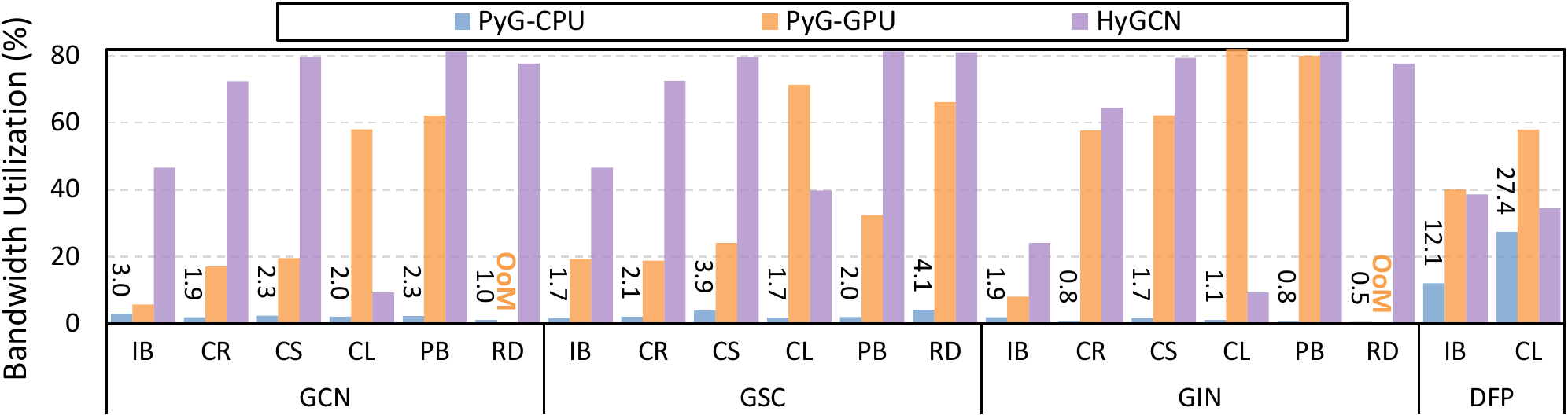}
    \vspace{-20pt}
    \caption{Bandwidth utilization of all platforms.}
    \label{fig:bandwidth_utilization}
    \end{minipage}\hfill
    \begin{minipage}{0.42\textwidth}
    \centering
    \includegraphics[page=1, width=\textwidth]{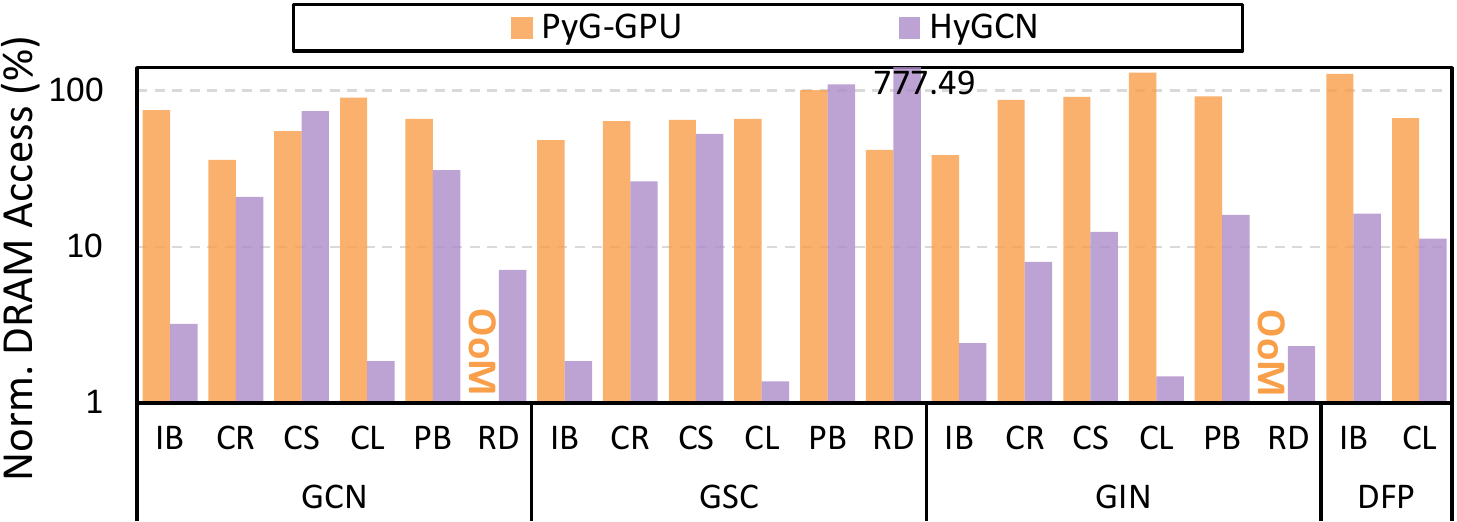}
    \vspace{-20pt}
    \caption{Normalized data access to PyG-CPU.}
    \label{fig:overall_data_access}
    \end{minipage}\hfill
\end{figure*}

\begin{figure*}[!htbp] 
    \centering
    \begin{minipage}{0.3\textwidth}
    \centering
    \vspace{3pt}
    \includegraphics[page=1, width=\textwidth]{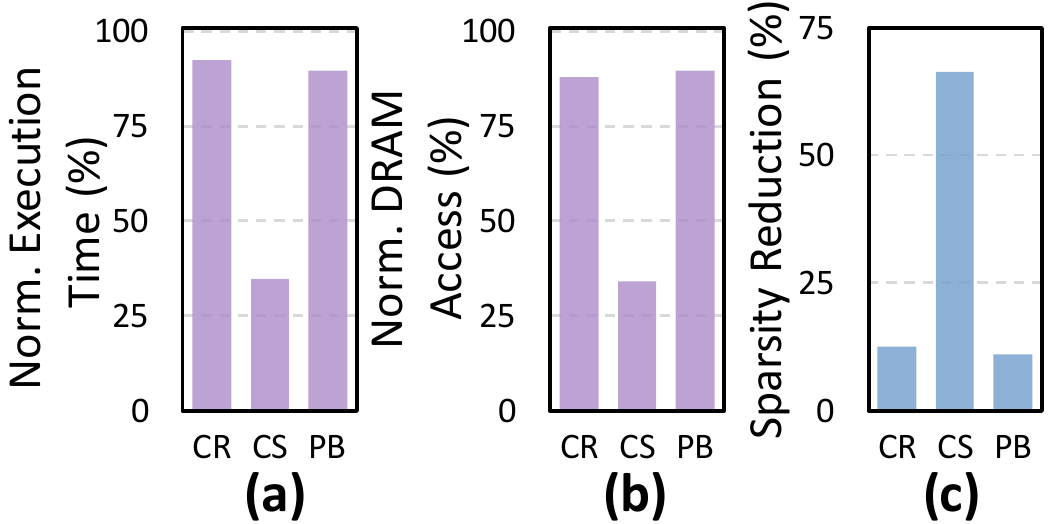}
    \vspace{-20pt}
    \caption{Effect of sparsity elimination on (a) execution time, (b) DRAM access, and (c) sparsity reduction.}
    \label{fig:sparsity_optimization}
    \end{minipage}\hfill
    \begin{minipage}{0.47\textwidth}
    \centering
    \includegraphics[page=1, width=\textwidth]{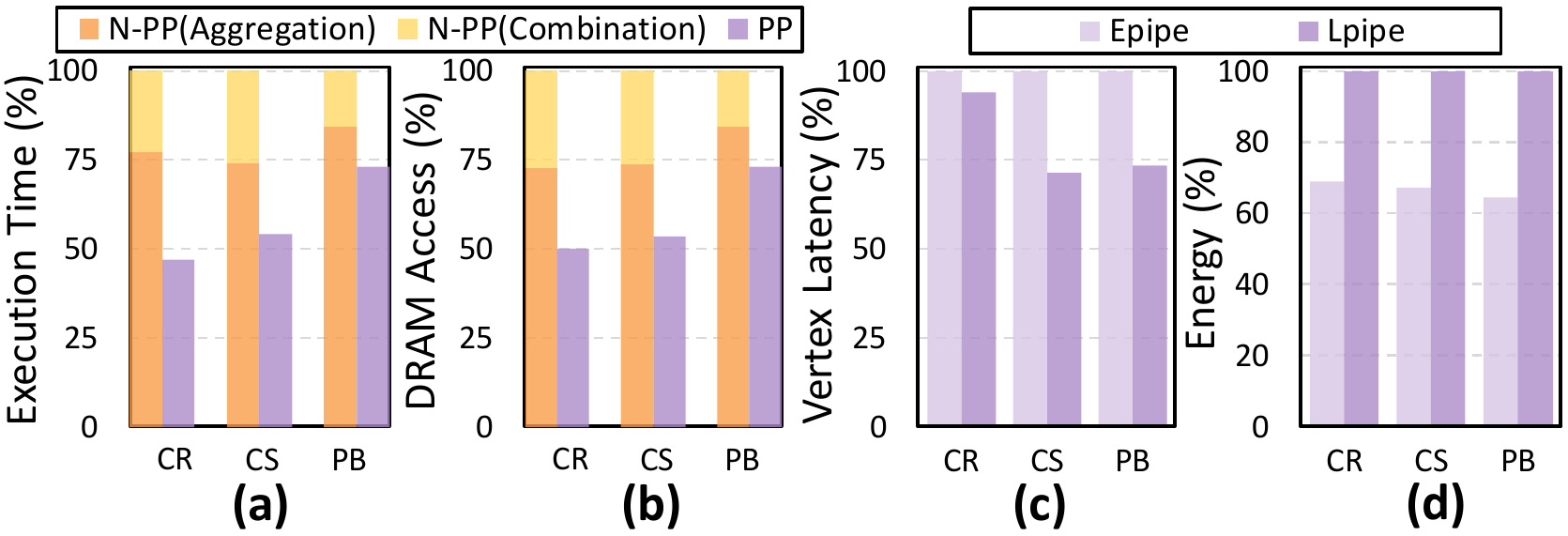}
    \vspace{-20pt}
    \caption{Effect of inter-engine pipeline on (a) execution time and (b) DRAM access, and the comparison of (c) vertex latency and (d) energy of \emph{Combination Engine} under different pipeline modes.}
    \label{fig:inter_engine_pipeline_optimization}
    \end{minipage}\hfill
    \begin{minipage}{0.20\textwidth}
    \centering
    \vspace{-3pt}
    \includegraphics[page=1, width=\textwidth]{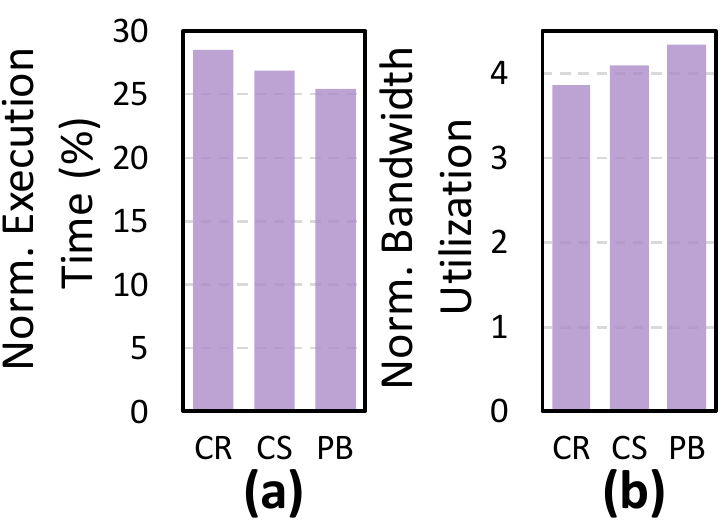}
    \vspace{-21pt}
    \caption{Effect of memory coordination on (a) execution time and (b) bandwidth utilization.}
    \label{fig:memory_coordination_optimization}
    \end{minipage}\hfill
    \vspace{-10pt}
\end{figure*}

As aforementioned, GIN causes additional computations and data accesses when performing \textit{Aggregation}, which introduces extra energy consumption on PyG-CPU. 
Although \emph{HyGCN} cannot reduce these computations, the optimizations of data reuse, sparsity elimination, and inter-engine pipeline can reduce redundant accesses to these additional data.
Among the architectural components, \emph{Combination Engine} consumes most of the energy due to the intensive computation of MVMs as depicted in Fig.~\ref{fig:energy_breakdown}, while \textit{Aggregation Engine} consumes more energy on high-degree graph datasets (i.e., CL and RD).

\noindent$\bullet$ \underline{\emph{\textbf{\textbf{DRAM Bandwidth Utilization.}}}} As seen in Fig. \ref{fig:bandwidth_utilization}, \emph{HyGCN} demonstrates 16$\times$ and 1.5$\times$ improvement on average on the utilization of DRAM bandwidth compared with PyG-CPU and PyG-GPU, respectively. 
The high bandwidth utilization of \emph{HyGCN} and PyG-GPU derive from the high-degree parallelism. 
By contrast, PyG-CPU cannot sufficiently exploit the bandwidth, since there is only one thread most of time to reduce the heavy overheads of frequent thread creation. 
Our consistent lower bandwidth on the CL dataset is due to the higher data reuse, which benefits from denser connections.

\noindent$\bullet$ \underline{\emph{\textbf{DRAM Access.}}} Although the 16MB on-chip memory is much smaller than the 60MB L3 cache on CPU and 34MB on GPU, \emph{HyGCN} accesses only 21\% and 33\% of off-chip data compared with PyG-CPU and PyG-GPU on average, respectively, as given in  Fig. \ref{fig:overall_data_access}. This benefits from our data reuse optimizations, sparsity elimination, and the immediate processing between two engines. On the CL dataset for GCN, GSC, and GIN, multiple graphs are assembled to form a larger one before being processed, which results in intensive sparsity. \emph{HyGCN} can efficiently eliminate the sparsity via window sliding and shrinking, thus avoiding unnecessary data accesses. Whereas, PyG-CPU and PyG-GPU produce many unnecessary accesses due to the irregularity in \textit{Aggregation} phase and without the fusion of phase-by-phase execution. 
As aforementioned, the results of PyG-CPU and PyG-GPU do not include the data access of the Sampling operation. For example, the Sampling access volume of GSC is 56.5GB on the RD dataset. In contrast, our work only accesses 28GB data, compared with the total 58GB in PyG-GPU.

\subsubsection{Power and Area}
The total power and area of \emph{HyGCN} are only 6.7 $W$ and 7.8 $mm^2$, respectively.
For the on-chip buffer, we use eDRAM to reduce both the area and energy consumption. 
For the computation precision, we use 32-bit fixed point that is enough to maintain the accuracy of GCN inference. 
Table \ref{table:acc_layout} provides area and power breakdown in terms of buffer, computation, and control. 
The computation resources of two engines consume most of power ($>$64\%) and area ($>$44\%) to perform the edge-centric aggregation and MVMs-based combination. The \emph{Coordinator} occupies $\sim$35\% of the total area since it has a large \emph{Aggregation Buffer}. The control overhead is small (only 1.2\% power and $<$0.45\% area) owing to the simple implementations of \emph{eSched}, \emph{Sampler}, \emph{Sparsity Eliminator}, \emph{vSched}, \emph{Coordinator}, and \emph{Memory Handler}.

\begin{table}[!htbp]
    \vspace{-10pt}
	\caption{Layout characteristics of \emph{HyGCN}}\label{table:acc_layout}
	\centering
	\renewcommand\arraystretch{1}
    \resizebox{0.43\textwidth}{!}{
    \begin{tabular}{cccc}
    \toprule
                 \textbf{Module}                       &      \textbf{Component}       & \textbf{Power} (\%) & \textbf{Area} (\%) \\ \midrule
    \multirow{3}{*}{\textbf{\emph{Aggregation Engine}}} & Buffer      & 2.37  &   5.41  \\
                                        & Computation & 3.85   &   1.43  \\
                                        & Control    & 0.48     &   0.18  \\ \midrule
    \multirow{3}{*}{\textbf{\emph{Combination Engine}}} & Buffer      & 14.4     &   15.13 \\
                                        & Computation & 60.52    &   42.96  \\
                                        & Control     & 0.31      &   0.07  \\ \midrule
    \multirow{2}{*}{\textbf{\emph{Coordinator}}}        & Buffer      & 17.66    &   34.64
     \\
                                        & Control     & 0.41    &   0.19  \\ \bottomrule
    
    \end{tabular}
    }
\end{table}

\subsection{Optimization Analysis} \label{optimization_analysis}
In this subsection, we analyze the effect of our optimization techniques including sparsity elimination, inter-engine pipeline, and off-chip memory access coordination. The benchmark model is GCN mentioned in Table \ref{table:gcn_model}.

\subsubsection{Sparsity Elimination Optimization}
We evaluate \emph{HyGCN} with and without sparsity elimination. This experiment runs only \emph{Aggregation Engine} to avoid the interference of other blocks. Fig. \ref{fig:sparsity_optimization}(a) shows that \emph{HyGCN} achieves 1.1$\sim$3$\times$ speedup with the optimization of sparsity elimination. The performance gain is due to fewer redundant DRAM accesses as reflected in Fig. \ref{fig:sparsity_optimization}(b), which benefits from eliminated sparsity as given in Fig.~\ref{fig:sparsity_optimization}(c).

\subsubsection{Inter-Engine Pipeline Optimization}

First, we measure the overall performance with and without inter-engine pipeline optimization (PP v.s. N-PP). With the pipeline optimization, the execution time of GCN is reduced by 27\%-53\%, as shown in Fig.~\ref{fig:inter_engine_pipeline_optimization}(a). On one hand, the \emph{Aggregation Engine} and \emph{Combination Engine} work in parallel with inter-engine pipeline. On the other hand, the DRAM accesses occupy most of the execution time (see Fig.~\ref{fig:inter_engine_pipeline_optimization}(b)), therefore the inter-engine pipeline helps improve the performance by decreasing DRAM accesses of the intermediate aggregation results between two engines. It is observed from Fig.~\ref{fig:inter_engine_pipeline_optimization}(b) that total DRAM accesses are significantly reduced to only 50\%-73\% with this pipeline optimization.

Second, we compare the vertex latency and energy of \emph{Combination Engine} with energy-aware pipeline and latency-aware pipeline (Epipe v.s. Lpipe). From Fig. \ref{fig:inter_engine_pipeline_optimization}(c), the Lpipe reduces the average latency for each vertex by 7\%-29\% via the immediate processing without waiting for the aggregation results of many vertices. By contrast, as shown in Fig. \ref{fig:inter_engine_pipeline_optimization}(d), the Epipe saves energy consumption by 35\% via assembling a large group of vertices to process together for reusing weight parameters aggressively. In practice, the application requirement determines the pipeline mode.

\subsubsection{Memory Coordination Optimization}
To show the effect of the memory access coordination, we present the execution time and bandwidth utilization with and without coordination in Fig. \ref{fig:memory_coordination_optimization}(a) and Fig. \ref{fig:memory_coordination_optimization}(b), respectively. With the memory access coordination for address continuity, the DRAM row buffers are better utilized and the channel-/bank-level parallelism is better exploited, which saves 73\% of time and improves 4$\times$ bandwidth on average.

\begin{figure*}[!hptb] 
    \centering
    \includegraphics[width=\linewidth]{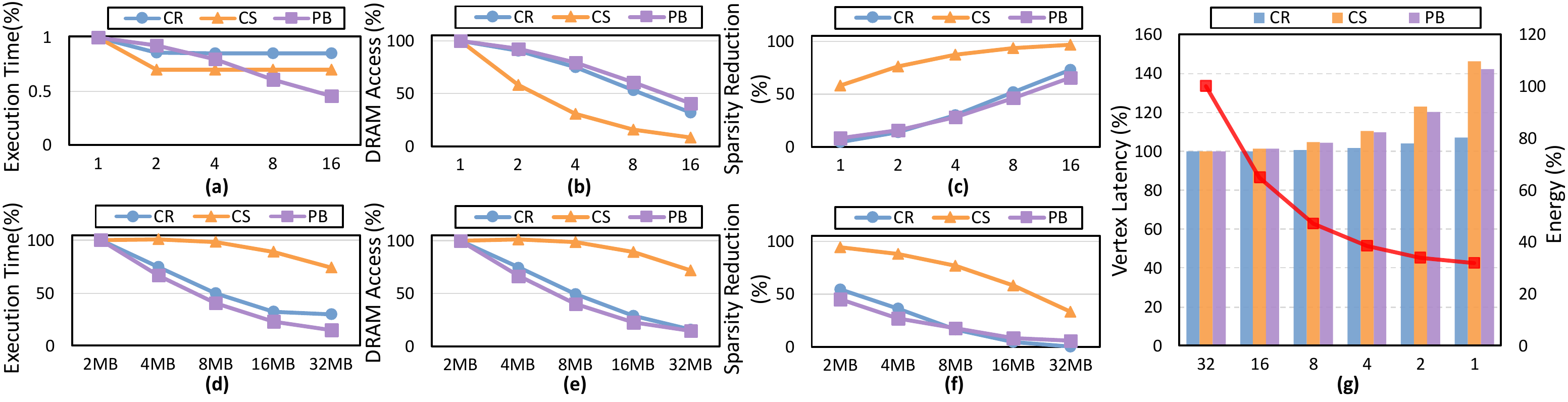}
    \vspace{-20pt}
    \caption{Scalability exploration. i) sparsity elimination with different \emph{sampling factor}: (a) Execution time, (b) DRAM access, and (c) sparsity reduction; ii) capacity of \emph{Aggregation Buffer}: (d) execution time, (e) DRAM access, and (f) sparsity reduction; iii) size of the systolic module: (g) vertex latency and energy of \emph{Combination Engine}.}
    \label{fig:scalability_exploration}
    \vspace{-10pt}
\end{figure*}

\subsection{Scalability Exploration} \label{scalability_exploration}
The following evaluations are measured in GSC model. 
\noindent $\bullet$ \underline{\emph{\textbf{Sparsity Elimination with Sampling.}}}
The Sampling operation increases the sparsity, thus it has the potential to enlarge the benefits produced by sparsity elimination. In Fig. \ref{fig:scalability_exploration}(a)-(c), horizontal axis sweeps the \emph{sampling factor}. It indicates that only $\frac{1}{sampling~factor}$ edges of each vertex are sampled to perform aggregation. As the increasing \emph{sampling factor}, the performance is significantly improved on the PB dataset by reducing the DRAM accesses owing to the higher sparsity. For other datasets, since many edges have been removed, the \textit{Combination} phase gradually dominates the execution time. Therefore, there is no significant speedup.
Note that the \emph{sampling factor} cannot be too high, as it might harm the accuracy of applications.

\noindent $\bullet$ \underline{\emph{\textbf{Capacity of Aggregation Buffer.}}}
The size of the \emph{Aggregation Buffer} affects the execution time, amount of data accesses, and even the effect of sparsity elimination. 
As the capacity of \emph{Aggregation Buffer} increases from 2 MB to 32 MB, the exeuction time is decreased as shown in Fig. \ref{fig:scalability_exploration}(d). 
This can be explained from two aspects: i) more intermediate aggregated feature data can be cached in on-chip buffer, leading to larger shard width when partitioning the graph and thus less execution loops; 
ii) larger shard means that the neighbor features can be reused more often, leading to less DRAM accesses (see Fig. \ref{fig:scalability_exploration}(e)). 
However, larger shard also enlarges the window size during the sparsity elimination, which results in higher sparsity that cannot be eliminated (see Fig. \ref{fig:scalability_exploration}(f)). 

\noindent $\bullet$ \underline{\emph{\textbf{Size of Systolic Module.}}}
In this experiment, we fix the number of total systolic arrays but change the size of each systolic module, and then to measure the cost of \emph{Combination Engine}. Different from the systolic module with 4$\times$128 systolic arrays in Table \ref{table:acc_system}, here we treat 1$\times$128 systolic arrays as a basic systolic module. Based on the initial 32 systolic modules, we gradually 
decrease the number of systolic modules under the restriction of fixed number of total systolic arrays. 
It is observed that longer latency for a vertex is consumed as the partition of systolic modules becomes more coarse-grained as shown Fig.~\ref{fig:scalability_exploration}(g)(bar). This is caused by the longer time to assemble a larger group of vertices to be processed together. Fortunately, the energy consumption can be reduced as shown Fig.~\ref{fig:scalability_exploration}(g)(red line) because the weight parameters are reused by more vertices within each larger systolic module. We only present the average energy result of these datasets for simplicity. In our architecture design, we set the systolic module with size of 4$\times$128 arrays to achieve a good trade-off between the latency and energy costs.

%% file: tex/discussion.tex
\section{Discussion}

In order to leverage our proposed PM, PyG needs to be significantly modified for its coarse-grain message-passing mechanism to stream \textit{Aggregation} and \textit{Combination} for each vertex. 
Note that although these two phases can be streamed after modification, it also misses the advantage of hardware-optimized operations, such as matrix multiplication operation~\cite{MKL,cuBLAS}. 
Furthermore, further challenges exist with 1) inefficient memory subsystem due to workload-agnosticism~\cite{Graphicionado}, 2) difficulty in data reuse like systolic arrays~\cite{TPU}, and 3) expensive on-line preprocessing for workload reorganization and streaming. 

Following concerns make training unsuitable as a starting work to explore GCN hardware. 
First, training involves three passes with data dependency: forward, backward, and update, whose compute and memory patterns are more complex than that of inference with only the forward pass. Second, the gradient propagation in graphs is far more complicated than layer-by-layer propagation in neural networks.
However, training accelerators can leverage our architecture to design the forward pass, and would need specialized blocks for other passes and an efficient memory hierarchy to connect them.

%% file: tex/related.tex
\section{Related Work}
Plenty of software frameworks for graph analytics and neural networks have been presented to release the programming efforts while achieving high performance on modern general-purpose architectures \cite{GraphMat,Gunrock,TensorFlow,PyTorch}. However, all of them only work well for the single-pattern workloads. Therefore, a large number of software frameworks for hybrid-pattern GCNs are proposed recently \cite{PyTorch_Geometric,pytorch_biggraph,DGL,web_scale_GCN,AliGraph}. 
For instance, PyTorch Geometric~\cite{PyTorch_Geometric} leverages message-passing framework to enhance its expression ability and the hardware-optimized operations (e.g. scatter and matrix multiplication) so that the GCN workloads can be accelerated.
Unfortunately, the distinct execution pattern regarding computation and access between the \emph{Aggregation} phase and the \emph{Combination} phase produces processing inefficiencies on traditional platforms. GCNs demand specialized architecture design.

With the emergence of graph analytics and neural networks workloads, a lot of hardware architecture designs are proposed to accelerate these workloads \cite{TPU,DianNao,eyeriss,Graphicionado,ozdal_energy_2016}. 
For example, Graphicionado~\cite{Graphicionado} is tailored for graph analtyics; while TPU \cite{TPU} focuses on the acceleration of neural networks.
However, GCNs behave not only like the graph processing (\emph{Aggregation}) but also like neural networks (\emph{Combination}), leading to intrinsic hybrid design requirement. Therefore, current specialized architectures cannot efficiently perform GCNs since they just handle one of the two sides.

%% file: tex/conclusion.tex
\section{Conclusion}

\emph{GCNs} are becoming widely adopted for analyzing graph data and are comprised of \emph{Aggregation} and \emph{Combination} phases. 
In this work, we identify that the execution patterns of these two phases are distinct, even almost opposite, which requires separate design requirements. 
Besides, the high intra-vertex parallelism in \textit{Aggregation} phase, the highly reusable inter-vertex data in \textit{Combination} phase, and the opportunity to fuse phase-by-phase execution introduced by the new features of GCNs need to be leveraged for better performance. 
To this end, we propose a GCN accelerator, \emph{HyGCN}, with hybrid architecture. First, we build edge- and MVM-centric programming model to exploit various parallelisms and enable hardware transparency. Next, we propose the hardware design with two efficient engines to optimize the two phases correspondingly. The latency- and energy-aware inter-engine pipelines are orchestrated to improve the overall latency and energy according to system needs. The off-chip memory accesses between the two engines are carefully coordinated to improve the efficiency. Finally, through comprehensive evaluations, \emph{HyGCN} demonstrates significant improvements compared to the software framework running on CPU and GPU. 
We believe our work will stimulate more attention on specialized hardware for increasingly important GCNs.